%% file: a_EH_intermittent_online_JRNL.tex
\documentclass[11pt,onecolumn,draftclsnofoot]{IEEEtran}

\def\myrootdir{/Users/ayca/Dropbox/DRAFTS/}

\def\bibdirM{\myrootdir/MY_BIB}
\def\bibdirMM{\myrootdir/MY_BIB}
\def\bibdirC{\myrootdir/MY_BIB}

\usepackage[utf8]{inputenc}
\usepackage{cite}

\usepackage[cmex10]{amsmath}
\usepackage{amssymb,latexsym}
\usepackage{epsfig}
\usepackage{caption}
\usepackage{color}
\usepackage{graphicx}
\usepackage{subcaption}
\usepackage{bbm}
\usepackage{pst-plot}
\usepackage{pstricks-add}
\usepackage{breqn}
\usepackage{algpseudocode}
\usepackage{algorithm}
\usepackage{psfrag}
\usepackage{mathtools}

\algnewcommand{\Initialize}[1]{%
  \State \textbf{Initialize:}
  \Statex \hspace*{\algorithmicindent}\parbox[t]{.8\linewidth}{\raggedright #1}
}

\newcommand{\err}{\ensuremath{\varepsilon}} 
\newcommand{\myerr}{\ensuremath{\varepsilon}}

\newcommand{\sdq}{\ensuremath{{Q_d}}} 
\newcommand{\seq}{\ensuremath{{Q_e}}} 
\newcommand{\sq}{\ensuremath{{Q}}} 

\newcommand{\nts}{\ensuremath{{N_T}}} 

\newcommand{\emax}{\ensuremath{{E_u}}} 

\newcommand{\newoperator}[3]{\newcommand*{#1}{\mathop{#2}#3}}

\newoperator{\tr}{\mathrm{tr}}{\nolimits}
\newoperator{\diag}{\mathrm{diag}}{\nolimits}
\newoperator{\rank}{\mathrm{rank}}{\nolimits}
\newoperator{\myperm}{\mathrm{perm}}{\nolimits}
\newcommand{\expectation}{\ensuremath{\mathbb{{E}}}} 
\newcommand{\prob}{\ensuremath{\mathbb{P}}}

\newoperator{\myvec}{\mathrm{vec}}{\nolimits}

\newcommand{\herm}{\mathrm{\dagger}}

\newtheorem{thm}{\bf{Theorem}}[section]

\newtheorem{lem}{\bf{Lemma}}[section]

\newtheorem{rem}{\bf{Remark}}[section]

\newenvironment{theorem}
{\par\noindent \thm \begin{itshape}\noindent}
{\end{itshape}}
\newenvironment{lemma}
{\par\noindent \lem \begin{itshape}\noindent}
{\end{itshape}}
\newenvironment{remark}
{\par\noindent \rem \begin{itshape}\noindent}
{\end{itshape}}

\begin{document}

\title{Performance Bounds for Remote Estimation under  Energy Harvesting Constraints}

 \author{Ay\c ca \"Oz\c celikkale, Tomas McKelvey, Mats Viberg
 \thanks{A.~\"Oz\c celikkale, T.~McKelvey and M.~Viberg are with Dept. of Signals and Systems, Chalmers University of Technology, Gothenburg, Sweden  e-mails: \{ayca.ozcelikkale, tomas.mckelvey, mats.viberg\}@chalmers.se. }
 }
\date{}
\maketitle

\begin{abstract}
 Remote estimation with an energy harvesting sensor with a limited data and energy buffer is considered. The sensor node observes an unknown Gaussian field and communicates its observations to a remote fusion center using the energy it harvested. The fusion center employs minimum mean-square error (MMSE) estimation to reconstruct the unknown field. The distortion minimization problem under the online scheme, where the sensor has access to only the statistical information for the future energy packets is considered.  We provide performance bounds on the achievable distortion under a slotted block transmission scheme, where at each transmission time slot, the data and the energy buffer are completely emptied. Our bounds provide insights to the trade-offs between the buffer sizes, the statistical properties of the energy harvesting process and the achievable distortion. In particular, these trade-offs illustrate the insensitivity of the performance to the buffer sizes for signals with low degree of freedom and  suggest performance improvements with increasing  buffer size for signals with relatively higher degree of freedom.  Depending only on the mean, variance and finite support of the energy arrival process, these results provide practical insights for the battery and buffer sizes for deployment in future energy harvesting wireless sensing systems. 
\end{abstract}

\section{Introduction}
With the increasing number of mobile devices and wireless sensing applications, such as in smart cities, and the ever increasing demand on energy resources,  efficient usage of resources in sensing and communication systems is an urgent priority. In this respect, energy harvesting (EH), where devices are equipped with capabilities to collect energy from renewable sources such as solar power, provides a promising approach.
Here EH capabilities not only enable efficient usage of energy sources but also offer enhanced mobility and prolonged network life-times \cite{gunduzStamatiouMichelusiZorzi_2014,gilbert_comparison_2008,vullers_2010}.
Feasibility of energy harvesting approaches have been investigated and favourable results are obtained for various different scenarios, including harvesting from solar energy, mechanical energy sources and radio-frequency (RF) energy \cite{vullers_2010, gilbert_comparison_2008,visserVullers_2013,gorlatovaWallwaterZussman_2013}.

In parallel to these promising developments, there has been a significant effort to understand the { information transfer capabilities} of  communication systems with EH capabilities.
In the case of energy harvesting from RF sources, the main challenge lies in designing the transmission strategies at the transmitters  whose transmissions are utilized as information transfer by some nodes and energy transfer by some other nodes \cite{ZhangHo_2013,huangLarsson_2013,ParkClerckx_2014,ozcelikkaleDuman_2015_TWC}. 
In the case of systems which utilize energy harvested from natural sources, such as solar power, the key issue is the { intermittent nature of the energy supply}. Hence the main challenge is to provide reliable and efficient operation in these systems even when  the energy supply is unreliable. In this work, we focus on this intermittent nature of EH sources and its effect on the performance of remote estimation systems.

\subsection{Prior Work}

Performance of  communication systems under intermittent energy sources have been studied  under a broad range of scenarios.
Here an important distinction is the one between the {offline optimization scheme} and the {online optimization scheme} \cite{gunduzStamatiouMichelusiZorzi_2014}. In the offline (or deterministic)  scheme, the profile of the harvested energy is assumed to be known non-causally. In contrast, in the online (or stochastic) scheme,  only statistical knowledge about the energy arrivals is assumed to be known at the time of design.
The offline optimization scheme is relatively well-studied, especially in terms of formulations that adopt communication rate as the performance metric.
 Analytical results exist for various scenarios, including  point-to-point channels \cite{OzelTutuncuogluYangUlukusYener_2011,TutuncuogluYener_2012}, broadcast channels \cite{antepliUysalErkal_2011} and multiple-access channels \cite{yangUlukus_2012mac}.

In contrast, the online scheme is  considered to be  less tractable analytically.
A typical numerical method here is the dynamic programming approach, which utilizes a search over a quantized state space. Unfortunately, this approach not only has high computational complexity, which limits its applicability in low-complexity EH sensors, but it also falls short of providing systematic insight into the effect of system parameters \cite{gunduzStamatiouMichelusiZorzi_2014}.
On the other hand,  results that directly provide analytical insight for the online scheme are available only for a limited number of scenarios  \cite{ozelUlukus_2012,DongFarniaOzgur_2015,SrivastavaKoksal_2013,kazerouniOzgur_2015}.
The capacity under intermittent energy arrivals is shown to be the same as the capacity of  a channel with an average power constraint equal to the average recharge rate for the infinite battery case \cite{ozelUlukus_2012}. Approximate characterizations  are provided  for the finite battery case \cite{DongFarniaOzgur_2015}.
Power allocations that are as close to average energy arrival rate as possible  and power allocations that decrease exponentially are found to be optimal for rate maximization scenarios for the infinite battery \cite{SrivastavaKoksal_2013} and the finite battery with Bernoulli arrivals scenarios \cite{kazerouniOzgur_2015}, respectively.
  Under a binary decision scheme, where at each time instant the sensor makes a decision to transmit or not,  threshold-based policies are proven to be optimal for remote estimation of Markov sources \cite{NayyarBasarTeneketzisVeeravalli_2013}.
 A learning theoretic approach  where statistical parameters of energy arrival and data arrival processes are learned by the transmitter over time is proposed and convergence of the performance of this scheme to the performance of the online scheme is proven in \cite{blascoGunduzDohler_2013}.
Here we contribute to the analysis of the online scheme by providing performance bounds for the remote estimation scenario. We will discuss our approach in Section~\ref{sec:intro:contributions}.

Establishing a  close connection with the estimation of the unknown physical field and in particular the degree of sparsity,  hence varying degrees of correlation of the unknown signal, is an important aspect of performance evaluation for sensing systems. 
Here sparsity, or equivalently the degree of freedom of a signal family, refers to the effective low dimensionality of the unknown signal \cite{ayca_unitaryIT2014}.  
In addition to providing a reasonable model for physical fields,  the sparsity of the signal can be utilized to compensate for the unreliable nature of the energy sources in an EH system.
Yet, for the EH systems, structural results that directly exploit the sparsity or the correlation characteristics are available only for a limited number of scenarios, such as  estimation of a single parameter \cite{nourianDeyAhlen_2015,knornDeyAhlenQuevedo_2015},  Markov sources \cite{NayyarBasarTeneketzisVeeravalli_2013,NourianLeongDey_2014,calvofullanaMatamorosAntonHaro_2015}, circularly wide-sense stationary signals with static correlation coefficient \cite{ayca_eusipco2016}, two correlated Gaussian variables \cite{GangulaGunduzGesbert_2015},  and i.i.d. Gaussian sources, as a result of the findings of, for instance, \cite{ozelUlukus_2012,DongFarniaOzgur_2015,SrivastavaKoksal_2013, kazerouniOzgur_2015, zhaoChenZhang_2015,OrhanGunduzErkip_2015}.  As we will further discuss in Section~\ref{sec:intro:contributions}, here we contribute to this aspect by providing performance bounds for the estimation error which depends on the sparsity of the unknown signal and statistical properties of the energy arrival process.

\subsection{Contributions}\label{sec:intro:contributions}

In this work, we consider an EH sensor which  observes an unknown correlated Gaussian field and communicates its observations to a remote fusion center using the energy it harvested. The fusion center employs MMSE estimation to reconstruct the unknown signal. We consider the problem under a limited data and energy buffer constraint using a slotted transmission scheme where, at each transmission time slot,  the data buffer and the battery are completely emptied. 
Motivated by the high complexity and the high energy cost of source and channel coding operations, we consider an amplify-and-forward strategy  as in\cite{zhaoChenZhang_2015,nourianDeyAhlen_2015,knornDeyAhlenQuevedo_2015}.
This approach is further motivated by the optimality of uncoded transmission for Gaussian sources over additive white Gaussian (AWGN) channels under the mean-square error metric \cite{GastparRimoldiVetterli_2003,gastpar_2008} and  the corresponding results for the energy harvesting systems in the asymptotic regime \cite{zhaoChenZhang_2015}.
We focus on the online scheme.  A preliminary version of this setup is considered in \cite{ayca_isit2016}, where the energy arrival process is restricted to be a Bernoulli process and the signal model is restricted to circularly wide-sense stationary signals.

Using random matrix theory and compressive sensing tools,  we provide performance bounds that reveal the trade-off between the buffer size, the statistical properties of the energy harvesting process and the achievable distortion. 
Consistent with the compressive sensing results, our bounds illustrate the insensitivity of the performance to the buffer size for signals with low degree of freedom, and  the possible performance gain due to increasing buffer sizes for signals with relatively higher degree of freedom. 
These results, which depend on the sparsity of the signal to be observed  and the first and  the second order statistical properties of the energy arrival process, provide insights into buffer and battery size choices for future wireless energy harvesting systems.


An important special case we consider is the case of circularly wide-sense stationary (c.w.s.s.)  signals, which are a finite-dimensional analog of wide-sense stationary signals \cite{neeser_proper_1993,GrayToeplitzReview}.  
In addition to the above block transmission scheme, we also consider the strategy of transmission of equidistant samples for the low-pass c.w.s.s. signals.   
The equidistant sample transmission scheme is motivated by the sampling theorems for c.w.s.s. signals \cite{ayca_unitaryIT2014} and the relationship between c.w.s.s. signals and wide-sense stationary signals due to asymptotic equivalence of circulant and Toeplitz matrices \cite{GrayToeplitzReview}.  
Our performance bounds suggest that for low-pass c.w.s.s. signals similar performance can be obtained by both strategies of  block transmission (i.e. spreading the energy as much as possible on  all samples in the buffer)  and sending only  equidistant samples with all the energy in the battery.  
We also  compare these results with their off-line counterparts under total energy constraints. 
Our online performance bounds are observed to be consistent with these off-line results and reveal the possible flexibility in the energy management for sensing of low-pass c.w.s.s. signals under energy harvesting constraints.

The rest of the paper is organized as follows. In Section~\ref{sec:sys}, the system model is described.   Our performance bounds are presented in Section~\ref{sec:perf}. 
We consider the case of c.w.s.s. signals in Section~\ref{sec:cwss}. 
In Section~\ref{sec:num},  numerical illustration of the bounds is provided. The paper is concluded in Section~\ref{sec:conc}.

{\it Notation:}
    The complex conjugate transpose of a matrix $A$  is denoted by $A^\herm$. The spectral norm of a matrix $A$ is denoted by $||A||$. Positive semi-definite (p.s.d.) ordering for Hermitian matrices  is denoted by $\succeq$. $I_n $ denotes the identity matrix with $I_n \in \mathbb{R}^{n \times n}$. The $l_2$ norm of a vector $a$ is denoted by $\|a\|$.  Statistical expectation is denoted by $\expectation[.]$. We denote the expectation with respect to signals involved with $\expectation_S[.]$ and the expectation with respect to the energy arrivals with $\expectation_E[.]$ for the sake of clarity when needed.

 \section{System Model} \label{sec:sys}

\subsection{Signal Model}\label{sec:sys:sig}
The aim of the remote estimation system is to estimate the unknown   complex proper zero mean  Gaussian field ${\bf{x}} =[x_1,\ldots, x_N] \in  \mathbb{C}^{N \times 1}$,   $ {\bf{x}} \sim \mathcal{CN}(0,{K_{\mathbf x}})$ with $K_{\mathbf x} =\expectation[\bf{x}  \bf{x}^\herm ]$. Here,  the covariance matrix $K_{\bf{x}}$ models the possible correlation of the field values in time. Let $s$ be the number of non-zero eigenvalues of $K_{\bf{x}}$, i.e. rank of $K_{\bf{x}}$. Let  $K_{\bf{x}} =U_{s} \Lambda_{x,s}  U_{s}^\dagger$ be the (reduced) eigenvalue decomposition (EVD) of   $K_{\bf{x}}$, where  $\Lambda_{x,s} \in  \mathbb{C}^{s \times s}$ is the diagonal matrix of non-zero eigenvalues and $U_{s} \in \mathbb{C}^{N \times s}$ is the sub-matrix of $U$ corresponding to non-zero eigenvalues with $U \in \mathbb{C}^{N \times N}$. Let $P_x=\tr[K_{\bf{x}}]=\tr[\Lambda_{x,s}]$.   We consider  $\Lambda_{x,s}$ of the form $\Lambda_{x,s}= \frac{P_x}{s} I_s$. 
Here $s$ gives the number of degrees of freedom (d.o.f.), i.e.  the sparsity level of the signal family.

We note that this model covers signal families with a wide range of correlation structures. In particular, signals with rank one correlation matrices where the signal components have a correlation coefficient of $1$,  and the i.i.d. signals with $K_{\bf{x}} = \Lambda_{x,n}=I_n$ are covered with this model.
This type of models have been used to represent signal families that have a low degree of freedom in various signal applications, for instance  as a sparse signal model in compressive sensing literature \cite{TulinoCaireVerduShamai_2013,ayca_unitaryIT2014}.

\subsection{Sensing and Communications to the Fusion Center}
At time epoch $t$, the sensor observes the field value at time $t$, i.e. $x_t$.  The observations are held in a  buffer of finite size  $\sdq$ before transmission.
The buffer contents at the end of transmission time slot $k$, i.e. at the end of time epoch $k \sdq $, is given by ${\bf{\bar x}_k} =[x_{(k-1) \sdq +1},x_{(k-1) \sdq + 2} \ldots  x_{(k-1) \sdq + \sdq} ]$.  For convenience, $\nts \!=\! N/\sdq$ is assumed to be an integer, where $\nts$ gives the number of transmission blocks.
Motivated by optimality of uncoded transmission for Gaussian sources over AWGN channels \cite{GastparRimoldiVetterli_2003,gastpar_2008,zhaoChenZhang_2015} and the high complexity and the high energy cost of source and channel coding operations, we consider an amplify-and-forward strategy,  similar to \cite{nourianDeyAhlen_2015,knornDeyAhlenQuevedo_2015,zhaoChenZhang_2015}.
Hence, as illustrated in Fig.~\ref{fig:ehsensor} and Fig.~\ref{fig:schedule}, at the end of transmission time slot $k$, the sensor transmits the data in its buffer to a fusion center using an amplify-and-forward block transmission strategy as follows
\begin{align}\label{y_k}
 {\bf{\bar y}_k }=  \sqrt{p_k} {\bf{\bar x}_k} +{\bf{\bar{w}}_k}, \quad \quad k =1,\ldots,\nts,
\end{align}
where $p_k$, ${\bf{\bar{w}}}_k$ and ${\bf{y}}_{k}$ denote the amplification factor, the effective channel noise and the received signal at the fusion center for transmission time slot $k$, respectively.
The channel noise ${\bf{w}} =[{\bf{\bar w}}_1,\ldots,{\bf{\bar w}}_\nts] \in \mathbb{C}^{N \times 1}$ is modeled as complex proper zero mean Gaussian ${\bf{w}}  \in \mathbb{C}^{N \times 1}$ $ \sim \mathcal{CN}(0,K_{\bf{w}})$ with $K_{{\bf{{w}}}} =\expectation[{\bf{{w}}} {\bf{{w}}}^\herm ] = \sigma_{{w}}^2 I_{N}$.

The above type of  block transmission scheme  allows us to study the effect of finite buffer size on estimation and facilitates connections with uniform power allocation strategies which are optimal for i.i.d. sources in the offline scheme \cite{ozelUlukus_2012}, and the power allocation strategies matching the average arrival rate of EH process  optimal for i.i.d. sources in the online scheme \cite{SrivastavaKoksal_2013}.  It is also supported by the fact that for devices with low power budgets, it is  more energy efficient to send relatively larger amount of data at each transmission \cite{tirronen_2013}.
%

 \input{psIntermittent}
%
 \input{ps_ehArrivals}

Let $\tau$ be the duration associated with one transmission slot.
The  energy used by the sensor for communications at slot $k$  can be written as follows:
\begin{align}\label{eqn:ehCost}
J_{k} \!=\!  \tau \expectation_S[  || \sqrt{p_k} {\bf{\bar x}_{k}} ||^2]   \!=\! \tau  \sum_{t=1}^{\sdq}  p_k  \sigma_{x_{(k-1) \sdq + t}}^2, 
\end{align}
where $\expectation_S[.]$ denotes the expectation with respect to the unknown signal $\bf{x}$. 
For convenience, we normalize the time duration as $\tau =1$ in the rest of the paper.
%

\subsection{Energy Constraints at the Sensor:}
We consider a battery-aided operation where the energy is stored at a battery and used in regular time intervals. Let the initial energy stored at the battery be $0$. 
At time $t$, an energy packet of $0 \leq E_t < \infty $ arrives to the sensor, where the harvested energy process is an i.i.d. discrete-time stochastic process with mean $\mu_E$ and variance  $\varrho_E$ and $E_t \leq  \emax < \infty$.
At the end of time slot $k$, the total energy  that have arrived to the battery at this time slot is given by ${\bar{E}_k}$ as follows
\begin{align}\label{eqn:barEk}
{\bar{E}_k}=\sum_{t=1}^{\seq} E_{(k-1) \seq + t}.
\end{align}
We assume that the transmission time slots for the data buffer and the battery is synchronized and $\seq=\sdq=\sq$.  We assume that the battery capacity $C$ satisfies $C \geq \emax Q$ so that  a total energy of $\emax Q$ can be stored at the battery.

In general, the sensor has to operate under the  energy neutrality conditions:
$
\sum_{l=1}^k  J_{l} \leq \sum_{l=1}^k  {\bar{E}_l},   k = 1,\ldots,\nts
$.
These conditions ensure that the energy used at any transmission time slot does not exceed the available energy.  Here, we focus on the case where at each transmission all the energy at the battery is used, i.e.
\begin{align}\label{eqn:ehneut}
J_k = {\bar{E}_k},  \quad \quad  k = 1,\ldots,\nts.
\end{align}
Here the left-hand side of \eqref{eqn:ehneut} depends on the power amplification factor $p_k$ at transmission time slot $k$ through \eqref{eqn:ehCost}. The right-hand side gives the available energy, i.e. the realization of the total energy stored at the battery at the end of transmission time slot $k$.
Performance of linear transmission strategies under such power constraints where the available energy  is modeled as a deterministic variable have been considered before, see for instance  \cite{BahceciKhandani_2008,LeePetersen_1976,TanGunduzVillardebo_2016} for formulations with total energy constraints and \cite{zhaoChenZhang_2015,nourianDeyAhlen_2015,knornDeyAhlenQuevedo_2015} for energy harvesting formulations. In this work, we consider this problem under stochastic energy harvesting constraints, i.e. the case where the energy available is modeled as a random variable.

\subsection{Estimation at the Fusion Center:}
 After $\nts$ transmission time slots, i.e.  obtaining $\mathbf{y} =[\mathbf{\bar y}_1,\ldots,\mathbf{\bar y}_\nts] \in \mathbb{C}^{N \times 1}$, the fusion center forms an estimate of $\bf{x}$. Let us consider a fixed $E_t$, $t=1,\ldots, N$ realization, hence $p_k$ are determined through \eqref{eqn:ehneut}.
The MMSE estimate conditioned on the energy arrivals $E_t$ can be found as \cite[Ch2]{b_andersonMoore_optFiltering}
\begin{align}\label{eqn:B:gen}
 \hat{\bf{x}} \!=\! \expectation_{S}[\mathbf{x} | \mathbf{y}] = K_{\mathbf{x y} } K_{\mathbf{y}}^{-1} \mathbf{y}.
\end{align}
Here  $\expectation_{S}[.]$ denotes  the statistical expectation  with respect to noise and signal statistics, including $\bf{x},\bf{w}$, but  not with respect to energy realizations (and hence not  with respect to $p_k$'s which are also a function of energy realizations).  This is the standard mean-square error estimator where the fusion center uses the source and noise statistics to form an estimate of the unknown variable \cite[Ch2]{b_andersonMoore_optFiltering}.
The resulting MMSE, $ \myerr = \expectation_S [ || \bf{x} -  \expectation[\mathbf{x} | \mathbf{y}] ||^2]$, can be expressed as follows   \cite[Ch2]{b_andersonMoore_optFiltering}
\begin{align} \label{err:invForm}
 \myerr=  \tr \left[( \frac {s}{P_x}I_s + \frac{1}{\sigma_w^2} U_s^\herm G  U_s)^{-1}\right],
\end{align}
where $G=\diag([{p_1} {\bf{1}}_\sdq,\ldots,{p_\nts} {\bf{1}}_\sdq])\in \mathbb{R}^{N \times N}$ and ${\bf{1}}_\sdq=[1,\ldots,1] \in \mathbb{R}^\sdq$ is the vector of ones. 
For the above standard MMSE estimation,  $p_k$'s are known at the fusion center. 
This is similar to the formulations using rate as the performance metric, where standard achievable rate expressions require full channel state and transmission power level information. 

In this work,  we are interested in performance bounds on the mean-square error $\myerr$ that hold with high probability with respect to energy arrivals. We present our bounds in Section~\ref{sec:perf} and Section~\ref{sec:cwss} and we provide numerical illustrations in Section~\ref{sec:num}.

\section{Performance Bounds}\label{sec:perf}	
We will now investigate the effect of different buffer sizes on the system performance.  We first provide our main results,  and then discuss  the performance of a related off-line scheme as benchmark.
Let us define
    \begin{align} \label{eqn:fbt}
f_{bt}(\mu,\varrho,r) &\doteq  2 s \exp\left(-\frac{\varrho}{\mu^2} h\left(\frac{\mu r}{\varrho} \right)\right)\\
\label{eqn:fbn}
f_{bn}(\mu,\varrho,r) &\doteq  2 s \exp\left(-\frac{r^2/2}{\mu r/3 + \varrho} \right)
 \end{align}
with $h(a)\doteq (1+a) \log(1+a)-a,\,\, a \geq 0$.

We now present our main result, i.e.  bounds on the error performance that hold with high probability:

 \begin{theorem}\label{thm:perf}
 Let $u_i \in \mathbb{C}^{s}$ denote the $i^{th}$ column of the matrix $U_{s}^\dagger$.
 Let $\eta_L =\min_i \| u_i \|^2$,  and $\eta_U =\max_i \| u_i \|^2$.
 Let $\emax$ be parametrized as $\emax= r_E \mu_E$, $r_E \geq 1$.
 The performance of the EH system satisfies the following bounds
 
I.
 \begin{align}\label{eqn:boundI}
   \prob( \myerr <  \err_I    ) \geq 1-f_{bt}\left(\mu_I, \varrho_I, r\right) \geq  1-f_{bn}\left(\mu_I, \varrho_I, r\right)
  \end{align}
  for  $r \in (0,\frac{1}{\eta_U}]$, where
         \begin{align}\label{eqn:boundIerr}
  \err_{I} &= \frac{1}{1+\frac{1}{\sigma_w^2} \mu_E (\frac{1}{\eta_U} -r)}  P_x \\
\label{eqn:boundImu}
    \mu_{I} &=    \frac{1}{\eta_L} \max \{ r_E-1, 1\}  \min\{\sq \eta_U, 	1\}\\
 \label{eqn:boundIvar}
    \varrho_{I} &= \frac{\varrho_E}{\mu_E^2} \frac{1}{ \eta_L^2} \frac{1}{Q}   \min\{Q \eta_U, 1\} 
   \end{align}

 II.
 
 \begin{align}\label{eqn:boundII}
   \prob( \err <  \err_{II}    ) & \geq 1-f_{bt}\left(\mu_{II}, \varrho_{II}, r\right) \\ 
   &\geq  1- f_{bn}\left(\mu_{II}, \varrho_{II}, r\right) 
  \end{align}
 for  $r \in (0,\frac{1}{\eta_U}]$,   where $\gamma \in [0,Q r_E]$ and 
   \begin{align}\label{eqn:boundIIerr}
  \err_{II} &= \frac{1}{1+\frac{1}{\sigma_w^2} \frac{1}{Q} \bar{p}  \gamma  \mu_E (\frac{1}{\eta_U} -r)}  P_x \\
  \label{eqn:boundIImu}
    \mu_{II}&= \frac{1}{\eta_L} \max \{ \frac{1}{\bar{p}}-1, 1\}  \min\{\sq \eta_U, 1\}  \\
 \label{eqn:boundIIvar}
    \varrho_{II}&=   \frac{1}{\eta_L^2}   (\frac{1}{\bar{p}}-1)  \min\{Q \eta_U, 1\} \\
\label{eqn:boundIIbarp}
  \bar{p} &= \prob(\bar{E}_k \geq   \gamma \mu_E) 
  \end{align}
  \end{theorem}

 {\bf{Proof:}} The proof is presented in Section~\ref{sec:app}.

For $\sq=1$, the energy $E_t$ that arrives to the sensor  at time $t$ is immediately used to send the sample $x_t$.  As the buffer size $\sq >1$ gets larger, the probability of sending the samples  in the buffer  (with non-zero power) increases since the probability of the battery being charged with nonzero energy also increases while waiting for the data buffer to be full. On the other hand, the power used to send each sample will be  lower compared to the case where the energy is used to send a fewer number of samples, for instance compared to the scenario of directly sending the sample $x_t$ with energy $E_t$ if an energy packet of  $E_t>0$ arrives ($Q\!=\!1$). Hence, the bounds here can be interpreted as an exploration of the trade-off between using a small number of samples with high signal-to-noise ratio (SNR), i.e. high power, and  a high number of samples with low SNR in the estimation process.

Here the scenario with $\sq=1$ is  closely related to  the classical compressive sensing setting.  In particular, let us consider the case where the energy arrival process can be modeled as an i.i.d.  Bernoulli random process.
A typical compressive sensing set-up is the scenario where the measurement process is modeled as an  i.i.d. process Bernoulli process where a measurement is made, for instance, when the Bernoulli random variable is $1$ and is dropped when the Bernoulli random variable is $0$.  Hence for $Q=1$,  the bounds presented here are closely related to the eigenvalue bounds provided in the compressive sensing literature \cite[Ch.12]{foucartRauhut_2013}.  In particular, for the scenario of $Q=1$ with static $\sigma_{x_t}^2 =\sigma_x^2$, (such as in the case of circularly wide-sense stationary signals) the bounds in Thm. 3.1 can be seen as a consequence of the eigenvalue bounds in the compressive sensing literature  \cite[Ch.12]{foucartRauhut_2013}, \cite{ayca_unitaryIT2014}.
For $\sq >1$ and non-uniform $\sigma_{x_t}^2$, Thm.~\ref{thm:perf} provides a set of novel eigenvalue bounds  for the formulation introduced in Section~\ref{sec:sys}.

In Section~\ref{sec:num}, we provide illustrations of these bounds. In the rest of this section, we compare these bounds with the performance of an off-line scheme. 

{\it{An off-line scheme with a total energy constraint: } }
To compare  the performance of our bounds in  Thm.~\ref{thm:perf},   we now consider an associated  off-line scheme.
In particular, we consider the case where the power amplification factors are not modeled as random variables that depend on the energy arrivals but deterministic variables to be optimized. 
Let us consider the case where each component $x_k$ is sent as follows: 
\begin{align}
y_t =  \sqrt{b_t} {x_t} +{w_t}, \quad \quad t=1,\ldots,N.
\end{align}
Here we introduced the notation $b_t \geq 0$ to denote the amplification factors to emphasize that these are modeled as deterministic variables as opposed to random variables.
We note that in contrast to the setting in Section~\ref{sec:sys}, here a block transmission constraint is not imposed onto the set of admissible sensor strategies (hence $Q=1$).
Let us denote the MMSE as follows: 
\begin{align} \label{err:invFormdet}
\bar{\err} (B) =  \tr \left[( \frac {s}{P_x}I_s + \frac{1}{\sigma_w^2} U_s^\herm B  U_s)^{-1}\right],
\end{align}
where $ b_t \geq 0, \forall t$  and  $B =\diag(b)=\diag([b_1,\dots, b_N]) \in \mathbb{R}^{N\times N}$. We focus on the following optimization problem
\begin{subequations}\label{eqn:offline-opt}
\begin{align}
   \err_d = \min_{\,\, \substack{B}} & \quad    \bar{\err}\left(B\right)   \\
\text{s.t.} \,\,
& \sum_{l=1}^N  b_{l}  \sigma_{x_{l}}^2 = E_{tot}, \quad \quad 
\end{align}
\end{subequations}
Here the sensor has a total energy of $E_{tot}$ and it can freely distribute this energy on the samples in order to minimize the error.
We note the following result: 
 \begin{lemma}\label{lem:offline}\cite[Lemma 3.3]{ayca_eusipco2016}
 An optimal strategy for \eqref{eqn:offline-opt} is given by uniform  $b_t$, i.e.  $B=\diag(E_{tot}/P_x)$ and the optimum value is given by 
 \begin{align}\label{eqn:offline}
  \err_{d}  = \frac{1}{1 + \frac{1}{\sigma_w^2} \frac{E_{tot}}{s} } P_x
 \end{align}
 \end{lemma}
\vspace{5pt}
We now consider the case where $E_{tot} = \mu_E N$, which corresponds to the total energy that will be obtained  if an energy packet of $\mu_E$  were harvested at each time step, i.e. the average energy of the energy arrival process. In this sense,   \eqref{eqn:offline} can be interpreted as a deterministic benchmark.  Hence \eqref{eqn:offline} becomes 
 \begin{align}\label{eqn:offline2}
  \err_{d}  = \frac{1}{1 + \frac{1}{\sigma_w^2} \mu_E  \frac{N}{s} } P_x
 \end{align}
 \begin{remark}
Comparing \eqref{eqn:offline2} and the error expressions in Thm.~\eqref{thm:perf} we observe that both expressions provide error expressions in the form $\frac{1}{1+\text{SNR}_{eff}} P_x$ where $\text{SNR}_{eff}$ takes the form $\text{SNR}_{eff}^D = \frac{1}{\sigma_w^2}\frac{N}{s} \mu_E$ for \eqref{eqn:offline2} and it takes the form  $\text{SNR}_{eff}^P = \frac{1}{\sigma_w^2} \mu_E  (1/\eta_U -r)$,  for instance,  for \eqref{eqn:boundIerr}.  Hence the error expressions in Thm.~\eqref{thm:perf} provide different operating points for how close one can operate to the deterministic scheme and with which probability through the variable $r$. We further explore this point in the context of circularly wide-sense stationary signals in the following section. 
\end{remark}

\section{Circularly Wide-Sense Stationary Signals}\label{sec:cwss}

In this section we specialize to the case of circularly wide-sense stationary signals, which constitute a finite dimensional analog of wide-sense stationary signals \cite{neeser_proper_1993,GrayToeplitzReview}.  The covariance matrix of c.w.s.s. signals is circulant by definition, i.e. the covariance matrix is determined by its first row as  $[K_{\mathbf{x}}]_{tk}=[K_1]_{\text{mod}_n (k-t)}$, where $K_1 \in \mathbb{C}^{1 \times n}$ is the first row of $K_{\mathbf{x}}$ \cite{neeser_proper_1993,GrayToeplitzReview}. 
%

The unitary matrix $U$ in the eigenvalue decomposition of the covariance matrices of c.w.s.s. signals is given by the Discrete Fourier Transform (DFT) matrix \cite{neeser_proper_1993,GrayToeplitzReview}.  Let $F^N$ denote the DFT matrix of size $N \times N$, i.e. $[F^N]_{tk} =(1/\sqrt{N}) \exp(-j \frac{2 \pi}{N} (t-1) (k-1))$, $1 \leq t,k \leq N$, where $j=\sqrt{-1}$.
Hence, the reduced EVD of $K_{\mathbf{x}}$ is given by $K_{\mathbf{x}} = F_\Omega^n \Lambda_{x,s} {F_\Omega^n}^\herm $, where $\Lambda_{x,s} = \diag(\lambda_k)=\frac{P_x}{s} I_s \in \mathbb{R}^{s \times s}$ and $F_\Omega^n \in \mathbb{C}^{n \times s}$ is the matrix that consists of  $s$ columns of $F^n$ corresponding to the non-zero eigenvalues.

Due to the circulant covariance matrix structure, the variances of the components of a c.w.s.s. signal satisfy $\sigma_{x_t}^2 = \sigma_x^2 = P_x/N, \forall t$.  Hence, $J_k = \sum_{t=1}^{\sdq}  p_k  \sigma_{x_{(k-1) \sdq + t}}^2 =p_k Q P_x/N$, and by \eqref{eqn:barEk}, \eqref{eqn:ehneut}, we have the following
\begin{align}\label{eqn:cwss}
p_k = \frac{N}{P_x}\frac{1}{Q}\sum_{t=1}^{Q} E_{(k-1) Q + t}.
\end{align}

For c.w.s.s. signals,  we have $\eta_L =\min_i \| u_i \|^2 =\frac{s}{N}$,  and $\eta_U =\max_i \| u_i \|^2 =\frac{s}{N}$ due to the DFT matrix.   Hence, \eqref{eqn:boundIerr}-\eqref{eqn:boundIvar} can be expressed as 
\begin{align}\label{eqn:boundI:etaconstant}
  \err_{I}&= \frac{1}{1+\frac{1}{\sigma_w^2} \mu_E \frac{N} {s}( 1-\tilde{r})}  P_x \\
\label{eqn:boundImu:etaconstant}
    \mu_{I} &=     \max \{ r_E-1, 1\}  \min\{\sq \frac{s}{N}, 	1\}\\
 \label{eqn:boundIvar:etaconstant}
    \varrho_{I} &= \frac{\varrho_E}{\mu_E^2}  \frac{1}{Q}   \min\{Q \frac{s}{N}, 1\} 
    \end{align}
for  $\tilde{r} \in (0,1]$.  Here we have scaled  $r, \mu_{I}, \varrho_{I}$ while going from  \eqref{eqn:boundIerr}-\eqref{eqn:boundIvar} to \eqref{eqn:boundI:etaconstant}-\eqref{eqn:boundIvar:etaconstant}, since $f_{bt}(.)$ and $f_{bn}(.)$ only depend on the ratios between $r, \mu_{I}, \varrho_{I}$.
Eqn. \eqref{eqn:boundIIerr}-\eqref{eqn:boundIIvar} can be specialized to the case of c.w.s.s. signals, similiarly. 
Hence, for c.w.s.s. signals \eqref{eqn:boundIerr} and \eqref{eqn:boundIIerr} can be expressed in terms an expression in the form of $\frac{1}{1+\text{SNR}_{eff}} P_x$, where  $\text{SNR}_{eff}$ becomes $ \text{SNR}_{eff} =\frac{1}{\sigma_w^2} \mu_E \frac{N}{s}( 1-\tilde{r})  $, $\tilde{r} \in (0,1)$  and $\text{SNR}_{eff}= \frac{1}{\sigma_w^2} \frac{1}{Q} \bar{p}  \gamma  \mu_E \frac{N}{s} ( 1-\tilde{r})  $, $\tilde{r} \in (0,1)$, for \eqref{eqn:boundIerr} and \eqref{eqn:boundIIerr} respectively.  
We note that the above performance bounds also hold for  other signal families for which  $\| u_i \|^2$ is constant for all $i$, such as unitary Hadamard matrices. 

{\it{ Comparison with the average performance of a greedy approach: }}
In the case of c.w.s.s. signals, the following bound on the  average error performance over different realizations of the process $E_t$ can be found:
 \begin{align}
 \label{eqn:erravgb1}
 \expectation_E[\err]  &\geq \tr \left[( \frac {s}{P_x}I_s + \frac{1}{\sigma_w^2} U_s^\herm \expectation [G]  U_s )^{-1} \right]\\
  \label{eqn:erravgb2}
 &= \tr [( \frac {s}{P_x}I_s + \frac{1}{\sigma_w^2}  U_s^\herm \diag( \frac{N}{P_x} \mu_E)  U_s )^{-1}] ] \\
\label{eqn:erravgb3}
& \geq \frac{1}{1 + \frac{1}{\sigma_w^2}\frac{N}{s} \mu_E } P_x
 \end{align}
 where $\expectation_E[.]$ denotes the expectation with respect to the energy arrival process.  In \eqref{eqn:erravgb1} we have used \eqref{err:invForm}, the Jensen's inequality and the fact that $\tr[X^{-1}]$ is convex for $X \succ 0$.
In  \eqref{eqn:erravgb2} and \eqref{eqn:erravgb3}, we have used $\expectation[G] \!=\!\expectation[p_k] I_N$, \eqref{eqn:cwss} and $U_s^\herm U_s \!=\! I_s$.  
We observe that this bound does not depend on $Q$. 
We note that  \eqref{eqn:erravgb3} is the same as the performance of  the off-line deterministic scheme in  \eqref{eqn:offline2}.  As in the case of \eqref{eqn:offline2}, the bounds of Thm.~\ref{thm:perf} can be  interpreted as a measure of how close one can operate  to \eqref{eqn:erravgb3}.

\subsection{Low-pass c.w.s.s. signals}
We now focus on the case of low-pass c.w.s.s. signals, i.e. c.w.s.s. signals for which the non-zero eigenvalues correspond to the low frequency components  $\Omega=\{0,1,s -1\}$. Such signals  can be recovered from their uniformly taken samples with zero mean-square error when the total number of samples is larger than the number of nonzero eigenvalues \cite{ayca_unitaryIT2014}. 
This property, which is consistent with the deterministic sampling theorems and the sampling theorems for wide-sense stationary signals \cite{Papoulis_1991book} motivates us to study strategies that send equidistant samples. The relationship between c.w.s.s. signals and wide-sense stationary signals due to asymptotic equivalence of circulant and Toeplitz matrices \cite{GrayToeplitzReview} further motivates this approach. 
In particular, we consider strategies that send one sample out of every $Q =N/s$ samples as follows: Let $t_d \in 0,\ldots, Q-1$, be  the fixed initial delay before sending the first sample and $N_T = N/Q \in \mathbb{Z}$ be the number of transmissions as before. 
Hence, we have 
$g_t \geq 0 $, if  $t= Q  (k-1) + t_d+1, \,\, 1 \leq k \leq N_T$, and $0$ otherwise. 
Hence,  the received signal at transmission time slot $k$ is a single sample as follows
\begin{align}\label{y_k:cwss}
y_k =  \sqrt{p_k} {x_{Q  (k-1) + t_d+1}} +w_k, 
\end{align} 
where $\sqrt{p_k}$ denotes the amplification factor, $w_k \in \mathbb{C}$,  $ w_k \sim \mathcal{CN}(0,\sigma_{\bar{w}}^2)$ denotes the i.i.d. complex proper Gaussian channel noise, as before.  The average energy used by the sensor for communication at slot $k$ can be written as follows: 
\begin{align}\label{eqn:ehCost:cwss}
J_{k} \!=\!   p_k  \sigma_{x_{(k-1) \sdq + t_d+1}}^2 =p_k \frac{P_x}{N},  
\end{align}
where we have used the fact that for c.w.s.s. signals $\sigma_{x_t}^2 = \sigma_x^2 = P_x/N, \forall t$.  By 
\eqref{eqn:ehneut}, we again have $J_k = {\bar{E}_k}$, $\forall k$.  We obtain the following bound for the performance of this system:

\begin{theorem}\label{thm:cwss}
The performance of the equidistant sample transmission strategy of \eqref{y_k:cwss} for low-pass c.w.s.s. signals  satisfies the following bounds:

 \begin{align}\label{eqn:boundI:cwss}
   \prob( \myerr <  \err_I^u   ) \geq 1-f_{bt}\left(\mu_I^u, \varrho_I^u, r^u\right) \geq  1-f_{bn}\left(\mu_I^u, \varrho_I^u, r^u\right) 
  \end{align}
  for  $r \in (0,1)$ where
         \begin{align}\label{eqn:boundI:err:cwss}
  \err_{I}^u &= \frac{1}{1+\frac{1}{\sigma_w^2} \mu_E \frac{N}{s} (1 -r)}  P_x \\
\label{eqn:boundImu:cwss}
    \mu_{I}^u  &=   \max \{ r_E-1, 1\}  \\
 \label{eqn:boundIvar:cwss}
    \varrho_{I}^u  &= \frac{\varrho_E}{\mu_E^2}  \frac{s}{N}  
   \end{align}

\end{theorem}

 {\bf{Proof:}} The proof is presented in Section~\ref{sec:pf:thm:cwss}.
 
 Comparing \eqref{eqn:boundI:err:cwss}-\eqref{eqn:boundIvar:cwss} with \eqref{eqn:boundI:etaconstant}-\eqref{eqn:boundIvar:etaconstant} for $Q=\frac{N}{s}$ reveals that for  low-pass c.w.s.s. signals, both the strategy of Thm.~\ref{thm:perf}, which spreads the energy accumulated  in the battery evenly on the samples in the buffer, and the equidistant sample transmission strategy of Thm. ~\ref{thm:cwss}, which uses the energy only on one sample from the buffer, provide the same performance bounds.  This property is consistent with the performance of the associated strategies in the  off-line scenario under a total energy constraint as discussed below:   %

{\it{An off-line scheme under equidistant sample transmission strategy: } }
Let us consider the equidistant sample transmission scheme in \eqref{y_k:cwss} under a total energy constraint as follows: 
\begin{subequations}\label{eqn:offline-opt:equidistantcwss}
\begin{align}
   \err_{de} = \min_{\,\, \substack{B_{u}}} & \quad    \bar{\err}\left(B_u\right)   \\
\text{s.t.} \,\,
& \sum_{l=1}^N  b_{l}  \sigma_{x_{l}}^2 = E_{tot}, \quad \quad 
\end{align}
\end{subequations}
under the condition  $b_t \geq 0$, if  $t= Q  (k-1) + t_d+1, \,\, 1 \leq k \leq N_T$, and $b_t=0$ otherwise; and $Q=N/s$,  $B_u =\diag(b_t)=\diag([b_1,\dots, b_N]) \in \mathbb{R}^{N\times N}$.%
\begin{lemma}\label{offline:equidistantcwss}
 An optimal strategy for \eqref{eqn:offline-opt:equidistantcwss} is given by $b_t =\frac{E_{tot}}{P_x}\frac{N}{s}$, if  $t= Q  (k-1) + t_d+1, \,\, 1 \leq k \leq N_T$ and $b_t=0$ otherwise. The optimum value is given by 
 \begin{align}\label{eqn:offline:equidistantcwss}
  \err_{de}  = \frac{1}{1 + \frac{1}{\sigma_w^2} \frac{E_{tot}}{s} } P_x
 \end{align}
\end{lemma}
{\bf{Proof:}}   We provide the proof  in Section~\ref{sec:pf:offline:equidistantcwss}. 

Hence, under the off-line scheme with a total energy constraint,  the performance of uniform power allocation over all samples, which is given by Lemma~\ref{lem:offline},  and the performance of the  equidistant sample transmission strategy given by  Lemma~\ref{offline:equidistantcwss} are the same.  Together with the fact that the performance bounds in the online case for block transmission scheme of Thm.~\ref{thm:perf} (specialized to c.w.s.s. signals in  \eqref{eqn:boundI:etaconstant}-\eqref{eqn:boundIvar:etaconstant}) and the performance bounds for the equidistant sample transmission scheme of Thm.~\ref{thm:cwss} are also  the same, this suggests flexibility in energy allocation for low-pass c.w.s.s signals in energy harvesting systems.

 \section{Numerical Results}\label{sec:num}

 \begin{figure}
\begin{center}
\psfrag{YYY}[bc][bc]{ MSE Bound }
\psfrag{XXX}[Bc][bc]{  {$p_{MSE}$} } 
\psfrag{DATADATA1}{\tiny  $\sq=1$}
\psfrag{DATADATA2}{\tiny $\sq=2$}
\psfrag{DATADATA3}{\tiny $\sq=4$}
\psfrag{DATADATA4}{\tiny$\sq=8$}
\includegraphics[width=0.8 \linewidth]{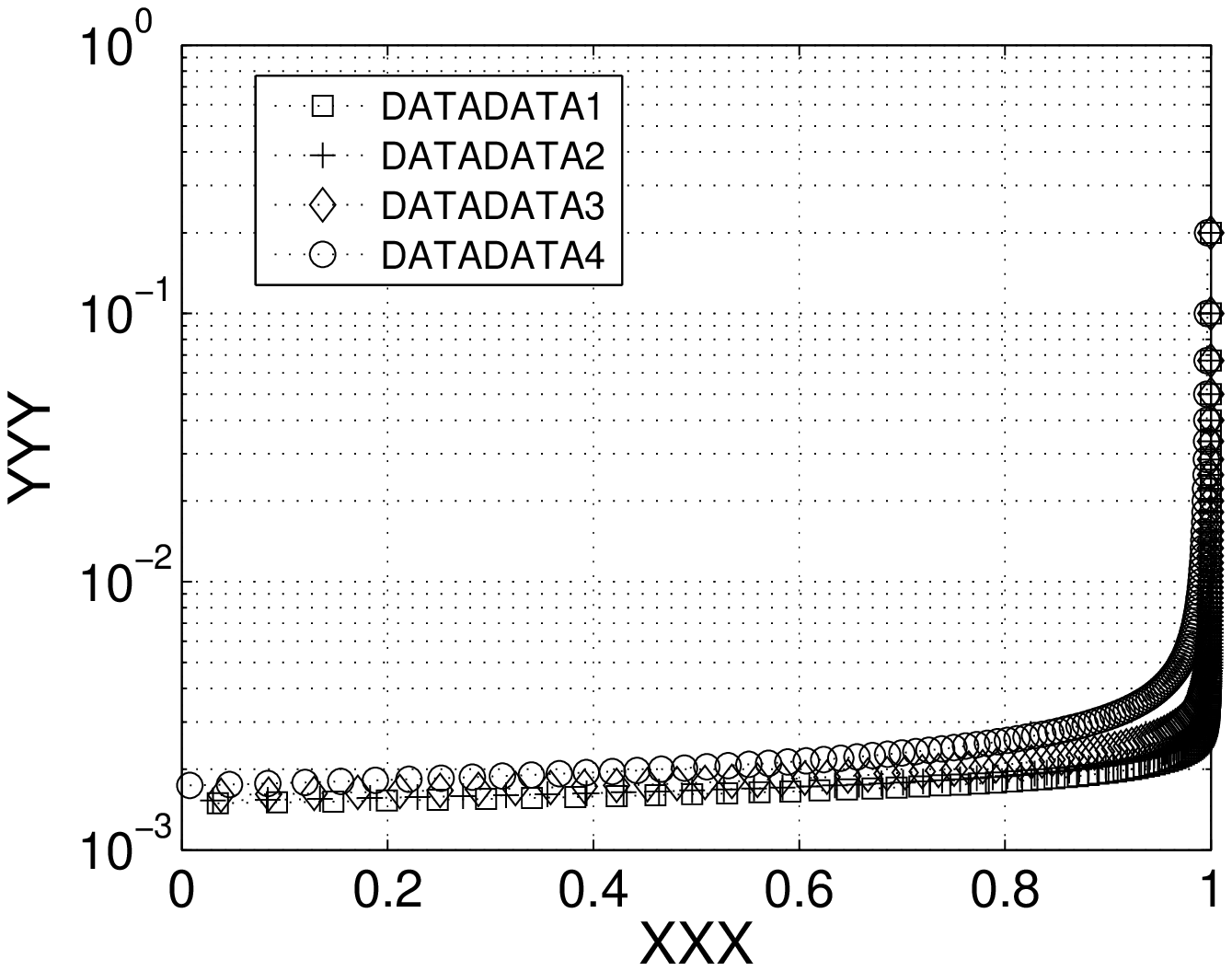}
\end{center}
\caption{MSE bound versus $p_{MSE}$, Bernoulli energy arrivals, $s = 4$.
}
\label{fig:eh1}
\end{figure}

\begin{figure}
\begin{center}
\psfrag{YYY}[bc][bc]{ MSE Bound }
\psfrag{XXX}[Bc][bc]{  {$p_{MSE}$} } 
\psfrag{DATADATA1}{\tiny  $\sq=1$}
\psfrag{DATADATA2}{\tiny $\sq=2$}
\psfrag{DATADATA3}{\tiny $\sq=4$}
\psfrag{DATADATA4}{\tiny$\sq=8$}
\includegraphics[width=0.8\linewidth]{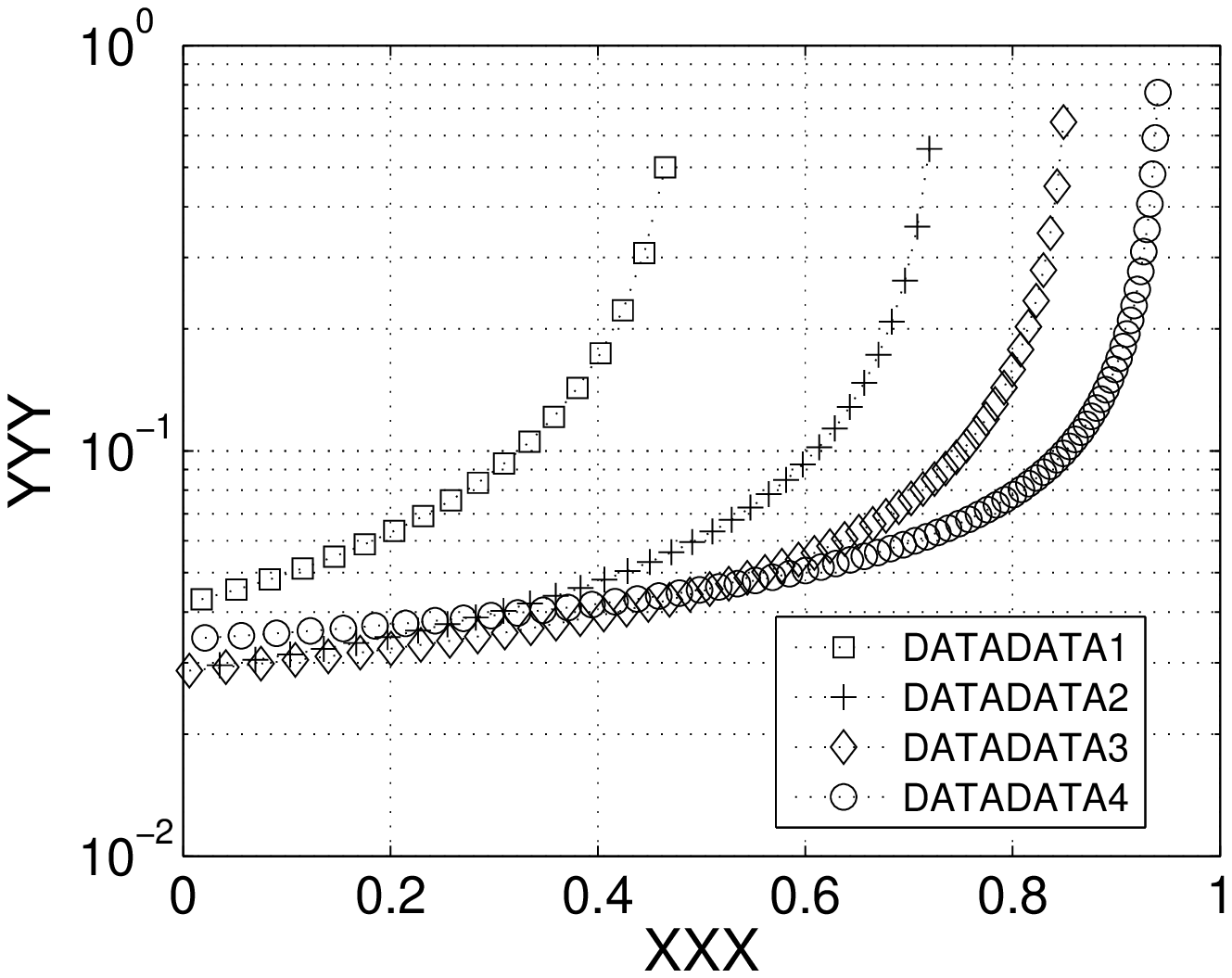}
\end{center}
\caption{MSE bound versus $p_{MSE}$, Bernoulli energy arrivals, $s = 16$.
}
\label{fig:eh2}
\end{figure}

We now illustrate the performance of our bounds. %
We first present the trade-off between the MSE bound and  the probability of obtaining that MSE performance.  Here, the y-axis corresponds to the error bound as provided by $\err_I$/$\err_{II}$ and the x-axis  corresponds to the probability on the right-hand side of \eqref{eqn:boundI}/\eqref{eqn:boundII}, which is referred as $p_{MSE}$.
While plotting the bounds, for a given probability value $p_{MSE}$ we present the tightest of Bound I and Bound II, i.e. the bound that provides the smallest error value with that given probability. 
We normalize the error bounds with the total uncertainty in the signal., i.e. we report $\err_I/P_x$  and $\err_{II}/P_x$. 

Let $N=256$, $P_x\!=\!N$, $\sigma_w^2\!=\!10^{-4}P_x$. Here we consider the energy arrival process with  $E_t$ i.i.d. with $E_t=\delta_t E_0$, $\delta_t\sim${Bernoulli} (p); $p\!=\!0.4$; $E_0=1$. 
We  consider the scenario of c.w.s.s. signals. 
The resulting bounds are presented in Figures ~\ref{fig:eh1} and ~\ref{fig:eh2}, for $s\!=\!4$ and $s\!=\!16$, respectively.  In both figures, as the desired performance becomes more demanding, i.e. the mean-square error (MSE) value decreases, the probability that this error can be guaranteed becomes smaller.
  When the degree of freedom of the signal is sufficiently low  ($s\!=\!4$, Fig.~\ref{fig:eh1}),  the performance bound  is observed to be relatively insensitive to the buffer size. On the other hand, when the degree of freedom is higher  ($s\!=\!16$, Fig.~\ref{fig:eh2}), the bound becomes more sensitive to the buffer size. For $s\!=\!16$, with small buffer sizes $Q\!=\!1,2$, the bound cannot provide any guarantees that hold with high probability; whereas with higher buffer sizes, small values of error can be guaranteed with high probability (for instance with probability higher than $0.8$ in Fig.~\ref{fig:eh2}).
 We note that here it is the Bound II that illustrates the behavior with $s\!=\!16$.
  We observe that as $s$ becomes larger, the signal can be said to be more close to an i.i.d. source, with the limiting case of $s\!=\!N$ corresponding to an exactly i.i.d source.    Hence these results are consistent with the results of  \cite{ozelUlukus_2012}, which show that for i.i.d. sources the strategies that spread energy on the samples as much as possible is an optimum strategy for the offline scheme. Here, the buffers and the slotted transmission scheme facilitate strategies  that are more close to such a uniform  allocation.

 \begin{figure}
\begin{center}
\psfrag{YYY}[bc][bc]{ MSE Bound }
\psfrag{XXX}[Bc][bc]{  {$p_{MSE}$} } 
\psfrag{DATADATA1}{\tiny  $\sq=1$}
\psfrag{DATADATA2}{\tiny $\sq=2$}
\psfrag{DATADATA3}{\tiny $\sq=4$}
\psfrag{DATADATA4}{\tiny$\sq=8$}
\includegraphics[width=0.76 \linewidth]{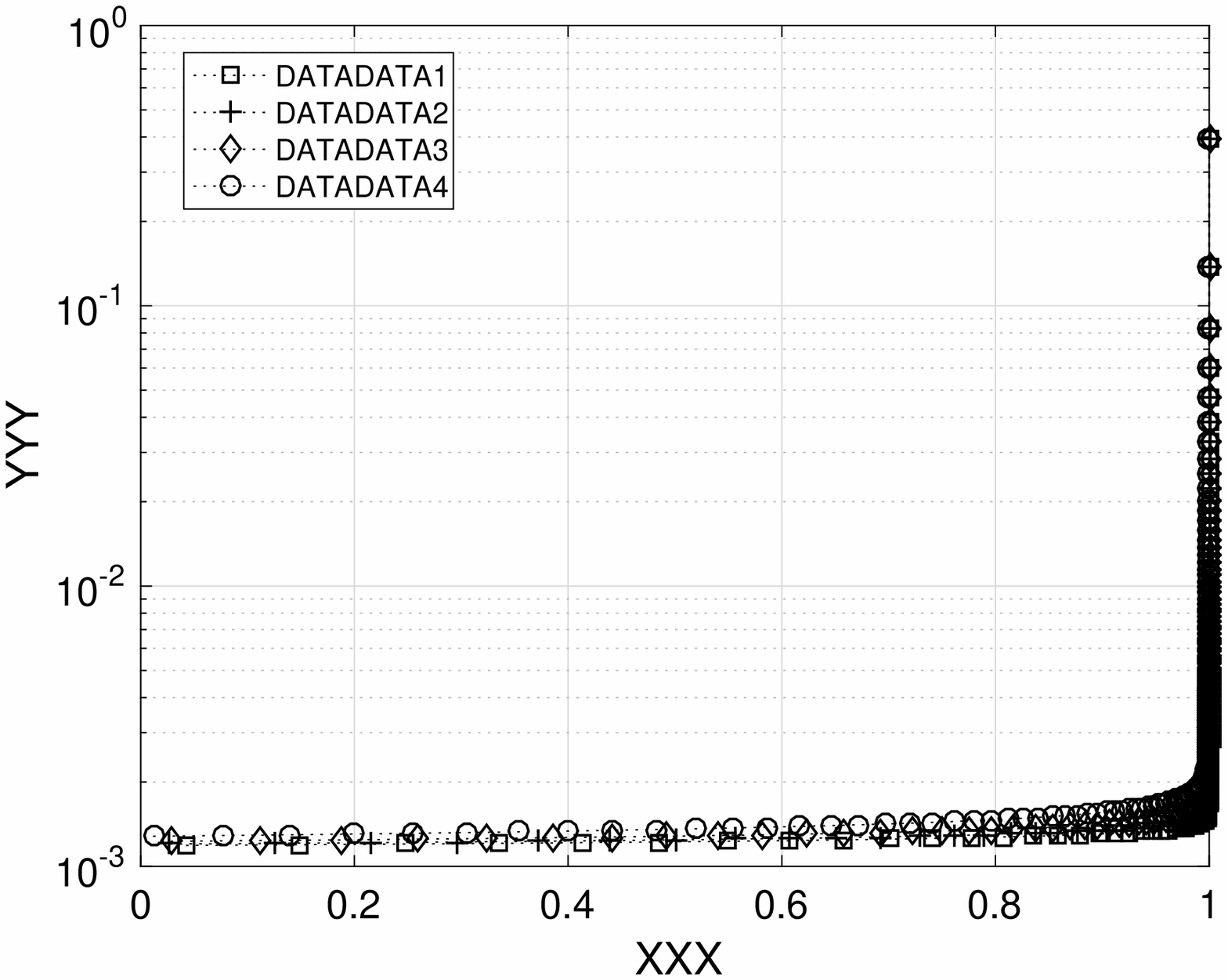}
\end{center}
\caption{MSE bound versus $p_{MSE}$, uniform energy arrivals,  $s = 4$.
}
\label{fig:eh1uniform}
\end{figure}

\begin{figure}
\begin{center}
\psfrag{YYY}[bc][bc]{ MSE Bound }
\psfrag{XXX}[Bc][bc]{  {$p_{MSE}$} } 
\psfrag{DATADATA1}{\tiny  $\sq=1$}
\psfrag{DATADATA2}{\tiny $\sq=2$}
\psfrag{DATADATA3}{\tiny $\sq=4$}
\psfrag{DATADATA4}{\tiny$\sq=8$}
\includegraphics[width=0.76\linewidth]{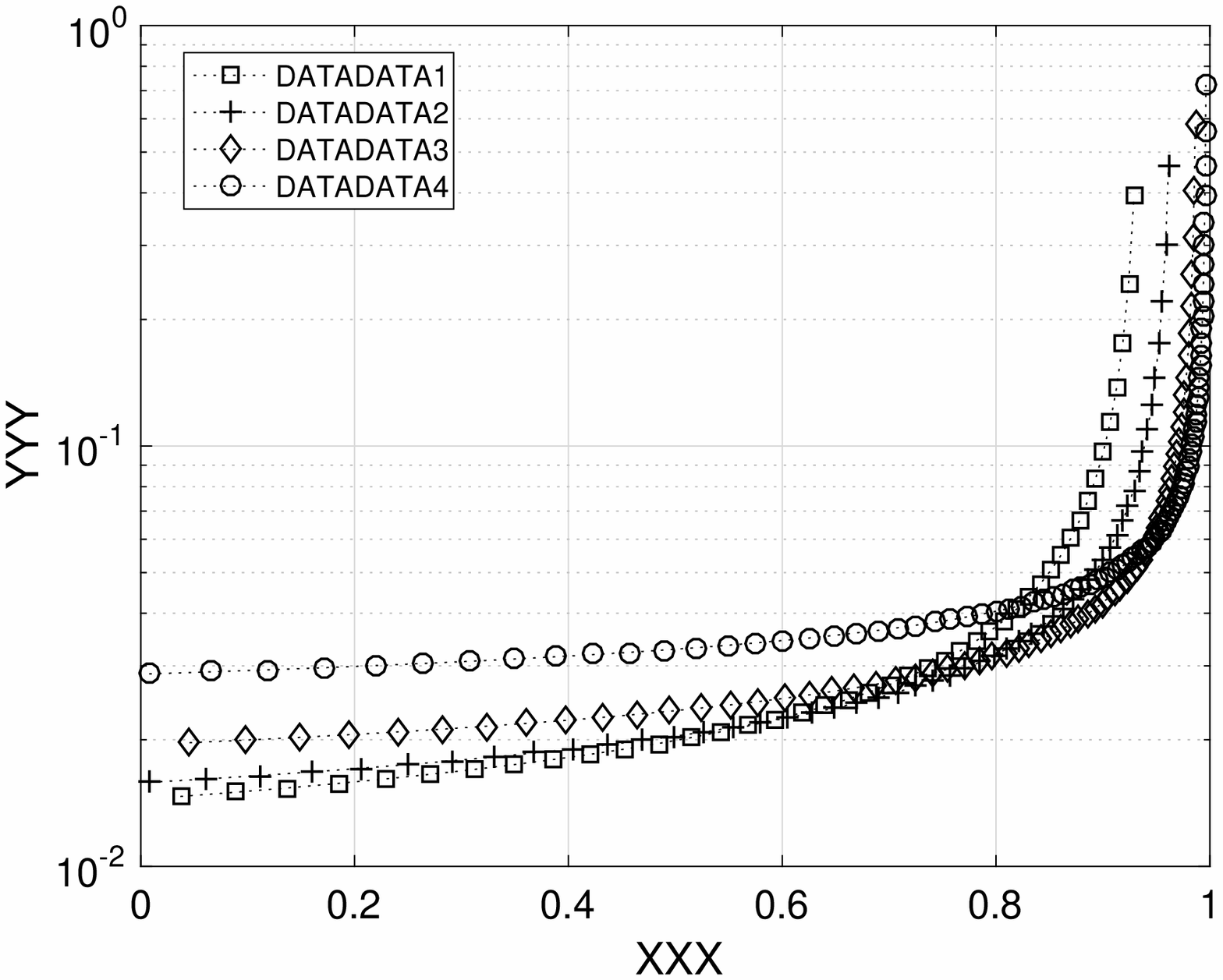}
\end{center}
\caption{MSE bound versus $p_{MSE}$, uniform energy arrivals,  $s = 16$.
}
\label{fig:eh2uniform}
\end{figure}

We now consider the above scenario with $E_t$ i.i.d. with uniformly distributed energy packets, i.e. $E_t\sim${Uniform}$[0, E_u]$; $E_u= 0.8$. Here $E_u$ is chosen so that the Bernoulli arrival case above and the uniform arrival case here has the same average energy.   The bounds for $s\!=\!4$ and $s\!=\!16$ are presented in Figures ~\ref{fig:eh1uniform} and ~\ref{fig:eh2uniform},  respectively. Consistent with Fig. ~\ref{fig:eh1}, the bounds are observed to be insensitive to the buffer size in the case of low degree of freedom, i.e.  $s\!=\!4$,  for  Fig. ~\ref{fig:eh1uniform}. 
Comparing Fig.~\ref{fig:eh2} and Fig.~\ref{fig:eh2uniform} for $p_{MSE} \geq 0.9$, we observe that  in the uniform energy arrival case of Fig.~\ref{fig:eh2uniform} long buffer lengths do not offer as much performance gain as they provide in the Bernoulli energy arrival case of Fig.~\ref{fig:eh2}. 
This is consistent with the fact that in the case of uniform arrivals the variance of energy packets is smaller which will decrease the need to spread the energy over samples by the use of a buffer. 

\begin{figure}
\begin{center}
\psfrag{YYY}[bc][bc]{\footnotesize MSE Bound }
\psfrag{XXX}[Bc][bc]{  \footnotesize{Error Probability} }
\psfrag{DATADATA1}{\tiny  $\sq=1$}
\psfrag{DATADATA2}{\tiny $\sq=2$}
\psfrag{DATADATA3}{\tiny $\sq=4$}
\psfrag{DATADATA4}{\tiny$\sq=8$}
\includegraphics[width=0.7\linewidth]{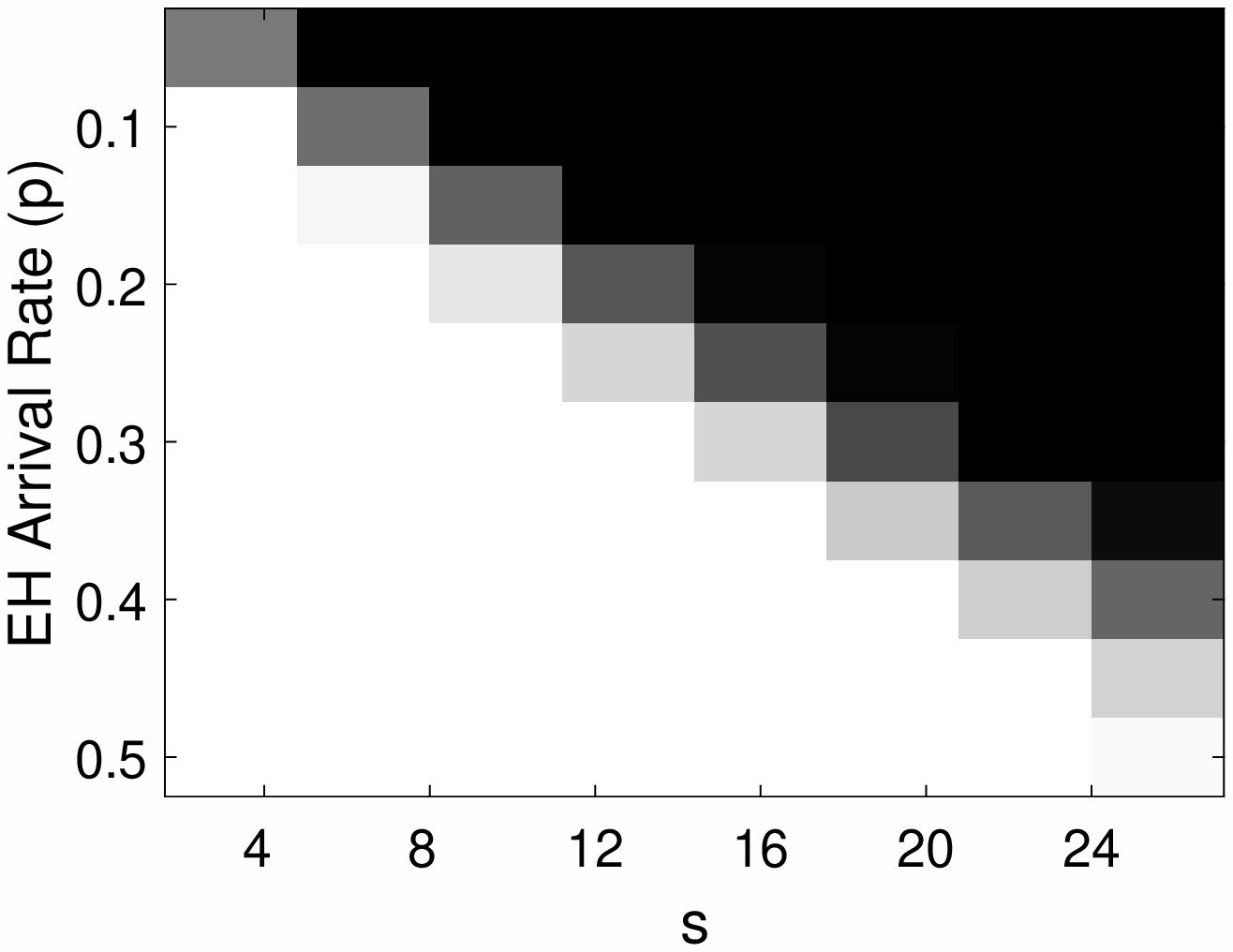} 
\end{center}
\caption{Empirical success probability as a function of energy arrival rate and $s$, $Q=1$.
}
\label{fig:sim1}
\end{figure}

\begin{figure}
\begin{center}
\psfrag{YYY}[bc][bc]{\footnotesize MSE Bound }
\psfrag{XXX}[Bc][bc]{  \footnotesize{Error Probability} }
\psfrag{DATADATA1}{\tiny  $\sq=1$}
\psfrag{DATADATA2}{\tiny $\sq=2$}
\psfrag{DATADATA3}{\tiny $\sq=4$}
\psfrag{DATADATA4}{\tiny$\sq=8$}
\includegraphics[width=0.7\linewidth]{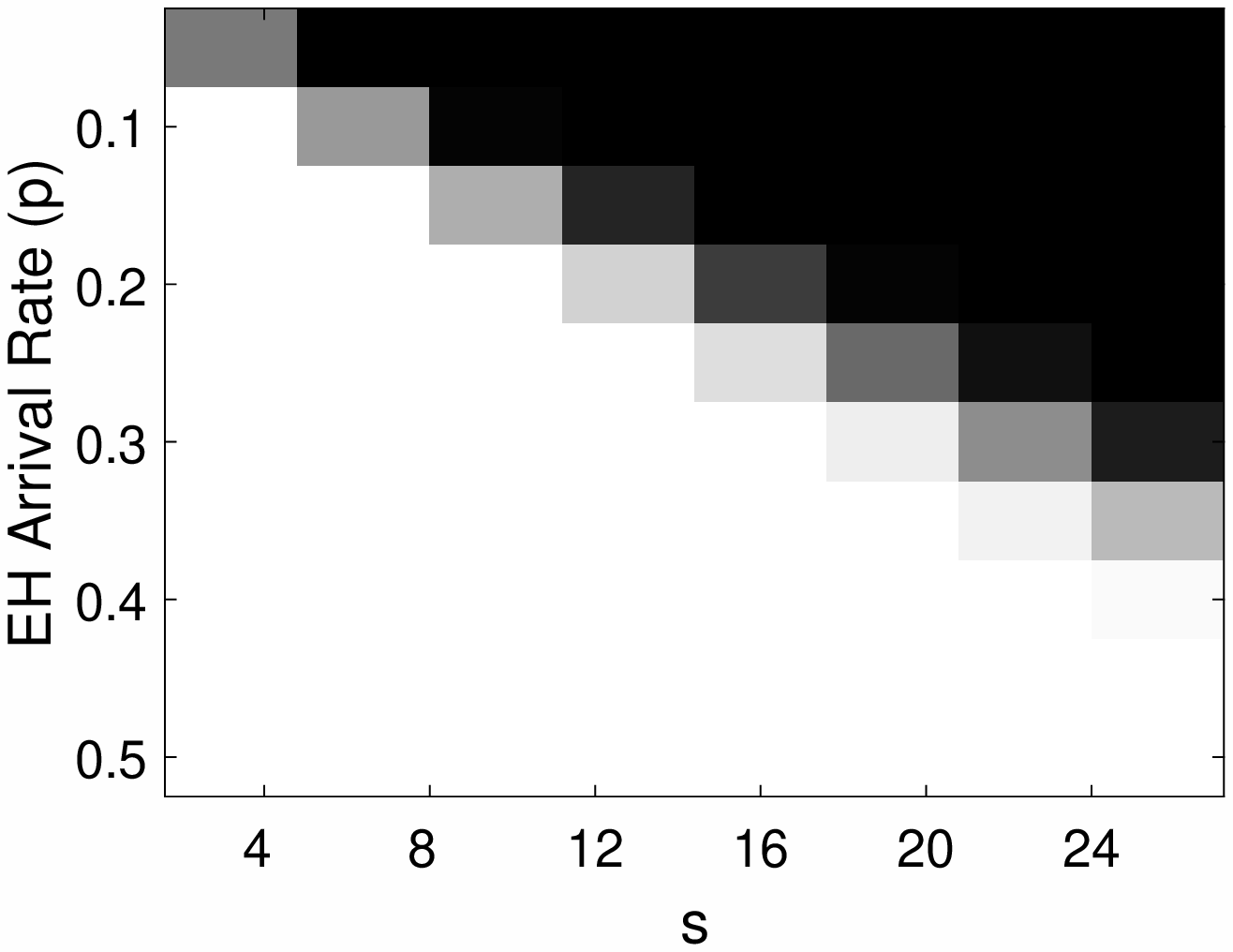} 
\end{center}
\caption{Empirical success probability as a function of energy arrival rate and $s$, $Q=32$.
}
\label{fig:sim2}
\end{figure}

We now provide numerical illustrations of the error performance.   We set an error threshold of $\err_{th}$. We simulate $N_{sim}$ realizations for the energy arrival process.  For each energy arrival realization, we calculate the MSE  $\err$ and define the event $\{ \err \leq  \err_{th}\}$ as a success. 
Let $\err_{th}=10^{-2}$, $N_{sim}=500$. Here $U$ is   the DFT  matrix and $E_t$ is  i.i.d. with $E_t=\delta_t E_0$, $\delta_t \sim${Bernoulli} (p); $E_0=1$. 
We report the empirical probability of success, i.e. ratio of successes to the total number of trials, as a function of the energy arrival rate $p$ and the sparsity level of the signal $s$ in Figures ~\ref{fig:sim1} and ~\ref{fig:sim2}, for $Q=1$ and $Q=16$ respectively.  
In the figures, lighter colours depict higher success rates with white depicting  $\%100$ success rate. 
We observe that the numerical results support the trends suggested by the performance bounds. 
In particular,  for signals with low degree of freedom (lower values of $s$) the buffer length does not affect the performance significantly. On the other hand for signals with relatively high degree of freedom (higher values of $s$),  longer buffer lengths enable better performance. 

 \kern-0.5em
\section{Conclusions}\label{sec:conc}
 We have considered remote estimation of a Gaussian field with an EH sensor with a limited data and energy buffer. 
 We have focused on the  stochastic energy harvesting framework where only statistical information about the online energy arrivals are available. 
 We have provided performance bounds on the achievable distortion under a slotted block transmission scheme. 
Our bounds provide insight into the trade-offs between the size of the buffer, statistical properties of the energy arrival process, the degree of freedom of the signal and the achievable distortion. 
These results also have the advantage of requiring only the knowledge of the mean,  variance and finite support about the energy arrival process, whose exact probability distribution can be difficult to reliably estimate in practice.

In particular, our results illustrate the insensitivity of the performance to the buffer size for signals with low degree of freedom as well as  the possible performance gain due to increasing buffer sizes for signals with relatively higher degree of freedom.  
Motivated by the sampling theorems,  we have also considered the scenario of equidistant sample transmission  for c.w.s.s. signals.  Our bounds for this scenario  suggest that similar performances can be obtained by both strategies of spreading the energy as much as possible on the samples in the buffer and sending only  equidistant samples with all the energy in the buffer.  These results, which are observed to be consistent with their off-line total energy constraint counterparts, suggest flexibility in sensing of low-pass c.w.s.s. signals with energy harvesting sensors.

\input{a_EH_intermittent_online_appJRNL}

\bibliographystyle{ieeetr}
\bibliography{\bibdirM/JNabrv,\bibdirM/bib_ayca,\bibdirMM/bib_randomMatrices,\bibdirM/bib_distributedEstimation,\bibdirM/bib_compressiveSampling2,\bibdirMM/bib_eh_M2M,\bibdirM/bib_energyHarvesting,\bibdirM/bib_csecuritySignal,\bibdirM/bib_robust,\bibdirMM/bib_robust_eh,\bibdirM/bib_optimization,\bibdirM/bib_books,\bibdirC/bib_ITrandomChannels,\bibdirMM/bib_mmse,\bibdirMM/bib_eh_intermittent,\bibdirMM/bib_eh_z,\bibdirMM/bib_mmse_comm,\bibdirMM/bib_fading,\bibdirMM/bib_hardware,\bibdirMM/bib_precoding_other,\bibdirMM/bib_eh_wsn,\bibdirM/bib_eh_prediction,\bibdirM/bib_linearEncoding}


\end{document}

%% file: psIntermittent.tex
\begin{figure}
\begin{center}
\begin{footnotesize}
\psset{arrowscale=1}
\psset{unit=0.6cm}
\psset{xunit=1.1,yunit=0.7}
\begin{pspicture}(-4,2.0)(6,7)
%
\psframe(0.7,2.5)(3.3,4)
\rput(2,3.2){{\small  EH Sensor}}
\rput(0.3,3.3){${ \mathbf{\bar{x}}_{k}}$}
\rput(4,3.2){${ \sqrt{p_k} \mathbf{\bar{x}}_{k}}$}
%
\psframe[linewidth=1pt,fillstyle=solid,fillcolor=lightgray](-1,3)(0,3.5)
\psframe[linewidth=1pt,fillstyle=solid,fillcolor=lightgray](-2,3)(-1,3.5)
\psline[linewidth=1pt,linecolor=black]{->}(-4,3.25)(-3,3.25)
\psline[linewidth=1pt](-3,3.45)(-2,3.45)
\psline[linewidth=1pt](-3,3.05)(-2,3.05)
\rput(-2,2.5){{\small Data} {\small Queue}}
\psframe[linewidth=1pt,fillstyle=solid,fillcolor=lightgray](1.75,5)(2.25,5.75)
\psline[linewidth=1pt,linecolor=black]{->}(2,7.0)(2,6.25)
\psframe[linewidth=1pt,fillstyle=solid,fillcolor=lightgray](1.75,6.0)(2.25,6.25)
\psline[linewidth=1pt](1.77,5.75)(1.77,6.5)
\psline[linewidth=1pt](2.22,5.75)(2.22,6.5)
\rput(3.5,6){{\small Battery}}
\rput(2,4.5){${J_k}$}
%
%
\psarcn[linewidth=0.4pt](4 ,3.2 ){1}{25}{-25}
\psarcn[linewidth=0.4pt](4 ,3.2 ){2}{30}{-30}
\end{pspicture}
\end{footnotesize}


\end{center}
\caption{ Energy Harvesting Sensor}
\label{fig:ehsensor}
\end{figure} 

%% file: ps_ehArrivals.tex
\begin{figure}
\begin{center}
\begin{tiny}
\psset{arrowscale=1}
\psset{unit=0.7cm}
\psset{xunit=0.9,yunit=0.8}
\begin{pspicture}(-2,0)(11,10)
\rput(-0.5,8.2){{\footnotesize Energy }}
\rput(-0.5,7.8){{\footnotesize Arrivals}}
\psline[linewidth=1pt,linecolor=black]{-}(1,8)(10,8)
\psdots[dotsize=3pt 3,linecolor=black,dotstyle=|](1,8)(2,8)(3,8)(4,8)(5,8)(6,8)(7,8)(8,8)(9,8)(10,8)
\psline[linewidth=1pt,linecolor=black]{->}(3.5,9.5)(3.5,8.5)
\rput(3.5,10){${\tiny E_Q}$}
\psline[linewidth=1pt,linecolor=black]{->}(4.5,9.5)(4.5,8.5)
\rput(4.5,10){${\tiny E_{Q+1}}$}
\psline[linewidth=1pt,linecolor=black]{->}(7.5,9.5)(7.5,8.5)
\rput(7.5,10){${\tiny E_t}$}
\rput(-0.5,5){{\footnotesize Data}}
\psline[linewidth=1pt,linecolor=black]{-}(1,5)(10,5)
\psdots[dotsize=3pt 3,linecolor=black,dotstyle=|](1,5)(2,5)(3,5)(4,5)(5,5)(6,5)(7,5)(8,5)(9,5)(10,5)
\psline[linewidth=1pt,linecolor=black]{->}(1.5,6.5)(1.5,5.5)
\rput(1.5,7){${\tiny x_1}$}
\rput(2.5,6.5){${ \dots}$}
\psline[linewidth=1pt,linecolor=black]{->}(3.5,6.5)(3.5,5.5)
\rput(3.5,7){${\tiny x_\sq}$}
\psline[linewidth=1pt,linecolor=black]{->}(4.5,6.5)(4.5,5.5)
\rput(4.5,7){${\tiny x_{\sq+1}}$}
\psline[linewidth=1pt,linecolor=black,linestyle=dotted]{-}(5.5,6.5)(8.5,6.5)
\psline[linewidth=1pt,linecolor=black]{->}(9.5,6.5)(9.5,5.5)
\rput(9.5,7){${\tiny x_{N}}$}
\rput(-0.5,1.5){{\footnotesize Transmission}}
\psbrace[rot=90,ref=t](1,4)(4,4){\tiny Transmission Time Slot $1$}
\psline[linewidth=1.5pt,linecolor=black]{->}(4,2.1)(4,1.2)
\rput(4,0.9){${\tiny  \sqrt{p_1} \mathbf{\bar{x}}_{1}}$}
%
\psline[linewidth=2pt,linecolor=black,linestyle=dotted]{-}(4.5,4)(6.5,4)
\psbrace[rot=90,ref=t](7,4)(10,4){\tiny Transmission Time Slot $\nts$}
\psline[linewidth=1.5pt,linecolor=black]{->}(10,2.1)(10,1.2)
\rput(10,0.9){${\tiny \sqrt{p_k} \mathbf{\bar{x}}_{\nts}}$}
\end{pspicture}
\end{tiny}
%

\end{center}
\caption{Time Schedule for the Energy Harvesting Sensor}
\label{fig:schedule}
\end{figure} 

%% file: a_EH_intermittent_online_appJRNL.tex

 \section{Proof of Thm.~\protect{\ref{thm:perf}}}\label{sec:app}

 We first prove the first family of bounds indexed by I in \eqref{eqn:boundI}-\eqref{eqn:boundIvar}.
 We first note that
\begin{align}\label{eqn:err:app}
 \err &=  \sum_{i=1}^{s} \frac{1}{\lambda_i(\frac{s}{P_x} I_s + \frac{1}{\sigma_w^2} U_s^\herm G  U_s)},\\
   \label{eqn:errboundsimple}
   &\leq   \frac{1}{1 + \frac{1}{\sigma_w^2}\frac{P_x}{s} \lambda_{min}(U_s^\herm  G  U_s)} P_x.
 \end{align}

In the remaining of the section, we let $k=1,\ldots,N_T$, $t=1,\ldots, \sq$ and use the indexing $z_{k,t}=z_{(k-1)Q+t}$ for any variable $z_i$, $i=1,\ldots,N$.
Let
\begin{align}\label{eqn:sumvar}
S_k  \triangleq \sum_{t=1}^{\sq} \sigma_{x_{k,t}}^2
\end{align}
where $\sigma_{x_{k,t}}^2 = \sigma_{x_{(k-1) \sq + t}}^2$.
By \eqref{eqn:ehneut} and \eqref{eqn:sumvar}, we have
\begin{align}
{p}_{k} = \frac{1}{S_k} \bar{E}_k  =\frac{1}{S_k} \sum_{l=1}^{\sq} E_{k,l},
\end{align}
where  $E_{k,l}= E_{(k-1) \sq + l}$.
Let $u_i \in \mathbb{C}^{s}$ denote the $i^{th}$ column of the matrix $U_{s}^\dagger$. Let $Y_{k,t} \triangleq   u_{k,t} u_{k,t}^\herm$, with $u_{k,t}= u_{(k-1) \sq + t}$.
Let us consider
\begin{align}
\bar{p}_{k} &\triangleq {p_{k}}-\expectation[p_{k}],\\
W_{k}& \triangleq \sum_{t=1}^{\sq}   Y_{k,t}\\
  \label{eqn:def}
 Z_k &\triangleq  \bar{p}_{k}  W_{k}
\end{align}
Hence
 \begin{align}
\sum_{k=1}^{\nts} Z_k 
& = \sum_{k=1}^{\nts} \sum_{t=1}^{\sq}  p_{k,t}   Y_{k,t} - \sum_{k=1}^{\nts}   \sum_{t=1}^{\sq} \expectation[p_{k,t}]   Y_{k,t},  \\
\label{eqn:Zidentity}
& =  U_s^\dagger G U_s - U_s^\dagger \bar{G} U_s,
\end{align}
where $\bar{G}=\diag([{\expectation[p_{1}]} {\bf{1}}_\sq,\ldots,{\expectation[p_\nts]} {\bf{1}}_\sq])\in \mathbb{R}^{N \times N}$ and ${\bf{1}}_\sq=[1,\ldots,1] \in \mathbb{R}^\sq$ is the vector of ones.
We will now use the Matrix Bernstein Inequality on $Z_k$ to find lower bounds for the eigenvalues of the first term in \eqref{eqn:Zidentity}. We will then use these in  \eqref{eqn:errboundsimple}  to bound the estimation error.

\vspace{4pt}
 \begin{lemma}[Matrix Bernstein Inequality \cite[Ch.8]{foucartRauhut_2013}]\label{lem:matrixBernstein} 
Let ${{V}}_1,\ldots,$ ${{V}}_M \in \mathbb{C}^s$  be independent zero-mean Hermitian random matrices. Assume that  $\|{{V}}_l \| \leq \mu_V$, $\forall l \in \{1,\ldots,M\}$ almost surely. Let
$ 
 \varrho_V \triangleq \| \sum_{l=1}^M \expectation[{{V}}_l^2]  \|.
$ 
Then, for $t>0$
\begin{align}\label{eqn:matrixBernstein}
 \prob( \| \sum_{l=1}^M {{V_l}}  \| \geq t ) \leq f_{bt}(\mu_V, \varrho_V,t) \leq f_{bn}(\mu_V, \varrho_V,t)
 \end{align}
with  $f_{bt}(.)$ and $f_{bn}(.)$ as defined in \eqref{eqn:fbt}-\eqref{eqn:fbn}.
\end{lemma}
\vspace{4pt}

%
We note that $Z_k$ in \eqref{eqn:def} are statistically independent random matrices with $\expectation[Z_k] ={0}$. We bound the spectral norm of  ${{Z}}_k$ as follows
\begin{align}\label{eqn:normbound0}
 \|{{Z}}_k \| =  \| \bar{p}_{k} W_k \|
             \leq  \left( \max_{k}  |\bar{p}_{k} | \right) \,   \|W_k\|.
\end{align}
%
We obtain the following bound for  $ \| W_k \|$
 \begin{align}\label{eqn:normbound1_1}
  \| W_k \| &= \| \sum_{t=1}^{\sq}  u_{k,t}  u_{k,t}^\herm \|, \\
  \label{eqn:normbound1_2}
   & \leq   Q  \max_{k,t} \|   u_{k,t}  u_{k,t}^\herm \|, \\
  & =  Q \max_{k,t} || u_{k,t}||^2,\\
      \label{eqn:normbound1}
   & = Q \eta_U,
\end{align}
where
\begin{align}\label{eqn:etaU}
\eta_U \triangleq \max_{k,t} || u_{k,t}||^2. 
\end{align}

We also have the following
 \begin{align} \label{eqn:normbound2_0}
   \|  W_k  \|
   \leq  \| \sum_{k=1}^{\nts}  W_k  \|
   =  \| I_s \| =1
\end{align}
where  \eqref{eqn:normbound2_0} follows from the fact that for $A \succeq 0$ and $B \succeq 0 $, $\lambda_{max}(A) \leq  \lambda_{max}(A+B)$. By \eqref{eqn:normbound1} and \eqref{eqn:normbound2_0}, we have the following
 \begin{align}\label{eqn:normboundf}
   \|W_k \| &\leq \min\{Q \eta_U , 1\}.
\end{align}

We now consider the term with $\bar{p}_{k} ={p_{k}}- {\expectation[p_{k}]} $ in \eqref{eqn:normbound0}
\begin{align}
 \label{eqn:normboundscalar0}
 \max_{k}  |{p_{k}}- {\expectation[p_{k}]}| &\leq  \max_{k}  \max \{  { p_{k}}-{\expectation[p_{k}]}, {\expectation[p_{k}]}\} \\
 &\leq  \max_{k}  \max \{  \frac{Q {\emax}-{Q\mu_E }}{S_k}, \frac{Q\mu_E}{S_k} \} \\
&\leq  Q  \max \{ \emax-\mu_E, \mu_E\} \frac{1}{\min S_k} \\
  \label{eqn:normboundscalar}
 &=  \mu_E \max \{ r_E -1, 1\}  \frac{1}{\eta_L} \frac{s}{P_x},
 \end{align}
 where we have used $\expectation[p_k] =Q \expectation[E_k] =Q \mu_E$, $p_k \leq  Q E_u $  and  $\emax =r_E \mu_E$. 
 Here \eqref{eqn:normboundscalar}  follows from 
 \begin{align}\label{eqn:SkL}
S_k =   \sum_{t=1}^{\sq} \sigma_{x_{k,t}}^2 \geq Q \min_{k,t}  \sigma_{x_{k,t}}^2 = Q \eta_L \frac{P_x}{s}, 
 \end{align}
where $ \sigma_{x_{k,t}}^2 = \frac{P_x}{s}  \| u_{k,t}  \|^2$  and 
\begin{align}\label{eqn:etaL}
\eta_L \triangleq  \min_{k,t} \| u_{k,t} \|^2
\end{align}

Hence by \eqref{eqn:normbound0}, \eqref{eqn:normboundf} and \eqref{eqn:normboundscalar}
\begin{align}\label{eqn:normbound}
 \|{{Z}}_k \| \leq     \mu_E \frac{s}{P_x} \frac{1}{\eta_L} \max \{ r_E-1, 1\}  \min\{\sq \eta_U, 	1\}   \triangleq \bar{\mu}_I, \quad \forall k.
\end{align}
We now consider the variance term, i.e.,
\begin{align}   \label{eqn:varbound0}
 \| \sum_{k=1}^{\nts} \expectation[ Z_k^2]  \| & =   \| \sum_{k=1}^{\nts} \expectation[ \bar{p}_k^2] W_{k}^2   \|,
  \\
 \label{eqn:varbound1}
 & \leq \max_k \expectation[ \bar{p}_k^2] \, \| \sum_{k=1}^{\nts} W_{k}^2   \|,
\end{align}
where we have used the fact that   $\expectation[ \bar{p}_k^2]  W_{k}^2  \preceq \! (\max_k \expectation[ \bar{p}_k^2]) W_{k}^2$ and  $\sum_{k=1}^{\nts} \expectation[ \bar{p}_k^2] W_{k}^2   \preceq    (\max_k \expectation[ \bar{p}_k^2]) \sum_{k=1}^{\nts} W_{k}^2  $. Here \eqref{eqn:varbound1} follows from  the fact that  for Hermitian $A$, $B$ with $A \succeq B$,  we have $\lambda_k(A)\geq \lambda_k(B)$, where  $\lambda_k(.)$ denote the ordered eigenvalues \cite[Cor. 7.7.4]{b_HornJohnson_TMA}. 
%
%

The spectral norm term in \eqref{eqn:varbound1} can be bounded as
\begin{align}
\label{eqn:var:spectral1}
\| \sum_{k=1}^{\nts} W_{k}^2   \| & \leq \max_k  \| W_{k} \| \| \sum_{k=1}^{\nts}  W_{k} \|,\\
\label{eqn:var:spectral2}
&\leq   \min\{Q \eta_U ,1\} 
\end{align}
where  \eqref{eqn:var:spectral1} follows from the fact that  $Z_k \succeq 0$, see for instance \cite[Sec. 2]{tropp_expected_2015}, and  \eqref{eqn:var:spectral2} follows from  \eqref{eqn:normboundf} and \eqref{eqn:normbound2_0}.

We now consider   $\expectation[ \bar{p}_k^2 ]$ in \eqref{eqn:varbound0}. We have the following
\begin{align}
\label{eqn:sumvariance0}
\expectation[ \bar{p}_k^2 ] &= \frac{1}{S_k^2} \sum_{l=1}^Q  \expectation[(E_{k,l} - \expectation[E_{k,l}])^2]\\
					&=  \frac{Q}{S_k^2}  \expectation[(E_{k,l} - \expectation[E_{k,l}])^2]\\\label{eqn:sumvariance2}
					&=  \frac{Q \varrho_E}{S_k^2}\\
\label{eqn:sumvariance3}
					& \leq \frac{\varrho_E}{Q \eta_L^2}(\frac{s}{P_x})^2,
\end{align}
where $\varrho_E$ is the variance of the energy arrival process as defined before,  \eqref{eqn:sumvariance0}  follows from the fact that $\bar{p}_k$ is a sum  of statistically independent zero mean variables and   \eqref{eqn:sumvariance3}  follows from $S_k^2 \geq Q^2 (\min_{k,t}  \sigma_{x_{k,t}}^2)^2 = Q^2 \eta_L^2 (\frac{P_x}{s})^2$.

Hence the variance term in \eqref{eqn:varbound0} can be bounded as follows
\begin{align} \label{eqn:varbound}
 \| \sum_{k=1}^{\nts} \expectation[ Z_k^2]  \| \leq   \frac{\varrho_E}{Q \eta_L^2} (\frac{s}{P_x})^2   \min\{Q \eta_U, 1\}  \triangleq  \bar{\varrho}_I.
\end{align}

Using \eqref{eqn:normbound}, \eqref{eqn:varbound} and the Matrix Bernstein Inequality reveals that for $\bar{r} >0$,
$
\| \sum_{k=1}^{\nts} Z_k \|   < \bar{r}
$
holds with probability greater than $p_{bt} = 1-f_{bt}(\bar{\mu}_I, \bar{\varrho}_I,\bar{r})$.
We note that for  Hermitian $A, B$, $\| A-B \| < \bar{r}$ implies $\lambda_{min}(A)> \lambda_{min}(B)-\bar{r}$. Therefore, using   \eqref{eqn:Zidentity},  with probability greater than $p_{bt}$ 
\begin{align}
\lambda_{min}(U_s^\dagger G U_s)
&> \lambda_{min}(U_s^\dagger \bar{G} U_s )-\bar{r}\\
\label{eq:eigdiff1}
&\geq  \min_{k} \expectation[p_{k}]-\bar{r} \\
\label{eq:eigdiff2}
&=\frac{\mu_E}{{\eta_U}} \frac{s }{P_x}-\bar{r} 
\end{align}
where \eqref{eq:eigdiff1} follows from the fact that $U_s^\dagger \bar{G} U_s =  \sum_{t=1}^{\sq} \expectation[p_{k,t}]   Y_{k,t} $ with $Y_{k,t} \succeq 0$ and
\begin{align}
 \sum_{t=1}^{\sq} \expectation[p_{k,t}]   Y_{k,t} \succeq \sum_{t=1}^{\sq} \min_{t,k} \expectation[p_{k,t}]   Y_{k,t} = \min_{t,k} \expectation[p_{k,t}] I_s
 \end{align}
 so that $U_s^\dagger \bar{G} U_s  \succeq   \min_{t,k} \expectation[p_{k,t}] I_s $.
Hence   $    \lambda_{min} (U_s^\dagger \bar{G} U_s) \geq    \min_{t,k} \expectation[p_{k,t}]$  due to the fact that  for Hermitian $A$, $B$ with $A \succeq B$,  we have $\lambda_l(A)\geq \lambda_l(B)$, where  $\lambda_l(.)$ denote the ordered eigenvalues \cite[Cor. 7.7.4]{b_HornJohnson_TMA}.
Here \eqref{eq:eigdiff2} follows from $\expectation[p_{k}] =  \frac{ Q \mu_E}{S_k} \geq \frac{\mu_E}{\eta_U}\frac{s}{P_x}$ where $S_k$  is bounded as follows
 \begin{align}\label{eqn:SkU}
S_k =   \sum_{t=1}^{\sq} \sigma_{x_{k,t}}^2 \leq Q \max_{k,t}  \sigma_{x_{k,t}}^2 = Q \eta_U \frac{P_x}{s}.
 \end{align}

Let us introduce $r$, $\mu_I$, $\varrho_I$, such that  $\bar{r} = \mu_E \frac{s }{P_x} r$, $\bar{\mu}_I =  \mu_E  \frac{s }{P_x} \mu_I$, $\bar{\varrho}_I= (\mu_E \frac{s }{P_x} )^2 \varrho_I$. Hence, \eqref{eq:eigdiff2} is expressed as
\begin{align}
\label{eqn:eigbound}
\lambda_{min}(U_s^\dagger G U_s)
&>   \mu_E  \frac{s}{P_x} (\frac{1}{{\eta_U}} -{r}).
\end{align}
We note that $\bar{\mu}_I, \bar{\varrho}_I,\bar{r}$ can be scaled as above without a change in the value of $f_{bt} (.)$ and $f_{bn} (.)$, i.e.  $f_{bt}(\bar{\mu}_I, \bar{\varrho}_I,\bar{r}) =f_{bt}({\mu}_I, {\varrho}_I,{r})$ and $f_{bn}(\bar{\mu}_I, \bar{\varrho}_I,\bar{r}) =f_{bn}({\mu}_I, {\varrho}_I,{r})$.
Using  $r$, $\mu_I$, $\varrho_I$, \eqref{eqn:eigbound}  and \eqref{eqn:errboundsimple} leads to the bounds in \eqref{eqn:boundI}-\eqref{eqn:boundIvar}. 

We now consider the second set of bounds given in \eqref{eqn:boundIIerr}-\eqref{eqn:boundIIbarp}.
 We first consider the event  $\bar{E}_k \geq   \gamma   \mu_E$   and define a new Bernoulli random variable $\bar{\delta}_k=\mathbbm{1}_{\bar{E}_k \geq   \gamma  \mu_E}$, where $\gamma \in [0,Q\, r_E)$  and $\mathbbm{1}$ is the indicator function.
We define  the probability $\bar{p}$ as follows
\begin{align}\label{eqn:barp1}
\bar{p}\triangleq \prob(\bar{E}_k \geq  \gamma \mu_E) 
\end{align}
Hence $ \prob(\bar{\delta_k}=1)=\bar{p}$ and $\prob(\bar{\delta_k}=0) = 1-\bar{p}$. Let us define 
\begin{align}\label{eqn:pkL}
{p}_k^L \triangleq \frac{ \gamma \mu_E }{S_k} \bar{\delta}_k.
\end{align}
We note that ${p}_k^L$ provides a lower bound for ${p}_k$, $\forall k$.
Hence we have $p_k W_k  \succeq {p}_k^L {W_k} $, $\forall k$,   and we have 
\begin{align}\label{eqn:succeq}
\sum_{k=1}^{\nts} p_k W_k  \succeq \sum_{k=1}^{\nts} {p}_k^L {W_k}. 
\end{align}
Hence the minimum eigenvalue of  $\sum_{k=1}^{\nts} {p}_k^L {W_k}$ provides a lower bound for the minimum eigenvalue of $\sum_{k=1}^{\nts} p_k W_k$  \cite[Cor. 7.7.4]{b_HornJohnson_TMA}.
Now  re-iterating the steps for the proof of bounds in \eqref{eqn:boundI}-\eqref{eqn:boundIvar} reveals a set of bounds similar to \eqref{eqn:boundI}-\eqref{eqn:boundIvar}, but that also depend on  $\gamma$. Here the variables related to ${p}_k$ are replaced with variables related to  ${p}_k^L$.  
In particular, $\bar{p}_k$ is replaced by $\bar{p}_k^L = p_k^L -\expectation[p_k^L] = p_k^L - \frac{\bar{p} \gamma  \mu_E}{S_k}$  and \eqref{eqn:normboundscalar} becomes $\bar{p}  \gamma  \mu_E \max \{\frac{1}{\bar{p}}-1,1\} \frac{1}{Q \eta_L} \frac{s}{P_x}$.  
Similarly,   \eqref{eqn:sumvariance3} becomes  $(\bar{p} \gamma \mu_E)^2 (\frac{1}{\bar{p}}-1) \frac{1}{Q^2 \eta_L^2} (\frac{s}{P_x})^2$ and  \eqref{eq:eigdiff2} becomes  $\frac{\bar{p} \gamma \mu_E}{{Q \eta_U}} \frac{s }{P_x}-\bar{r} $. 
Using these values,  and normalizing $\bar{\mu}_I, \bar{\varrho}_I,\bar{r}$  appropriately as before, we arrive at the bounds in  \eqref{eqn:boundII}-\eqref{eqn:boundIIbarp}.

\section{Proof of Thm.~\protect{\ref{thm:cwss}}}\label{sec:pf:thm:cwss}
We recall that $U_s = F_\Omega^N$  consists of  the first $s$ columns of the size $N$ DFT matrix  $F^N$.
The proof relies on the fact that equidistantly row sampled $F_\Omega^N$ can be associated with the DFT matrix of size $s$, $F^s$. 
Let $f_N=\exp(-j \frac{2 \pi}{N})$. The  entries of the row-sampled $F_\Omega^N$  can be expressed in terms of the entries of  $F^s$ as follows
\begin{align}
 \label{eqn:dftsampled0}
 [F_{\Omega}^N]_{(N/s)r+t_d+1, k+1} &=(1/\sqrt{N}) f_N^{((N/s)r+t_d) k} \\
 &=(1/\sqrt{N}) f_s^{r k} f_N^ {t_d k} \\
 \label{eqn:dftsampled}
 &= \sqrt{s/N} [F^s]_{r+1,k+1} f_N^ {t_d k},
\end{align} where  $ 0\leq k\leq s-1$, $ 0\leq r\leq s-1$.  
Now we adopt the arguments similar to Section~\ref{sec:app}.
By \eqref{eqn:ehneut} and \eqref{eqn:ehCost:cwss},  we have $p_k = \bar{E}_k  \frac{N}{P_x}$, where $\expectation[p_k] = Q   \mu_E  \frac{N}{P_x}$. 
Due to the equidistant sample transmission setting, we set  $W_k$  as follows
\begin{align}
W_k \triangleq Y_{(k-1)Q+t_d+1}, 
\end{align}
where $Y_{(k-1)Q+t_d+1}= Y_{k,t_d+1}$ and $Y_{k,t} \triangleq   u_{k,t} u_{k,t}^\herm$ as in Section~\ref{sec:app}.
Hence with $Z_k \triangleq (p_k -\expectation[p_k]) W_k$, we have
 \begin{align}
 \nonumber
\sum_{k=1}^{\nts} Z_k 
& = \sum_{k=1}^{\nts} p_k Y_{(k-1)Q+t_d+1} - \sum_{k=1}^{\nts}   \expectation[p_{k}]  Y_{(k-1)Q+t_d+1},  \\
\label{eqn:Zidentity:cwss}
& =  U_s^\dagger G U_s - U_s^\dagger \bar{G} U_s,
\end{align}
where  ${G}=\diag({g}_t) \in \mathbb{R}^{N \times N}$, $\bar{G} =\expectation[G]  \in \mathbb{R}^{N \times N}$ with ${g}_t =p_k $, if  $t= Q  (k-1) + t_d+1, \,\, 1 \leq k \leq N_T$, and $g_t = 0$ otherwise.

We note that $\eta_U$ and $\eta_L $  as defined in \eqref{eqn:etaU} and \eqref {eqn:etaL} is given by $\eta_U= \eta_L  =\frac{s}{N}$  due to  the fact that $U_s= F_\Omega^N$.
Hence we have 
$
  \| W_k \| =   \| u_{k,{t_d+1}} u_{k,{t_d+1}}^\herm \|  \leq  \eta_U,
$
with $\eta_U =s/N$. 
Due to \eqref{eqn:dftsampled}, we also have 
$
   \|  W_k  \|
   \leq  \| \sum_{k=1}^{\nts}  W_k  \| =   \| \frac{s}{N} F^s {F^s}^\dagger  \|
   =  \| \frac{s}{N} I_s \| =\frac{s}{N}
$.
Hence we bound    $ \|  W_k  \| $ as  $ \|  W_k  \| \leq \frac{s}{N}$.

Similar to \eqref{eqn:normboundscalar}, we also have  $ \max_{k}  |{p_{k}}- {\expectation[p_{k}]}| \leq   Q \mu_E \max \{ r_E -1, 1\}   \frac{N}{P_x} $, where we have used ${\eta_L} = s/N$.  
Hence we have 
\begin{align}\label{eqn:normbound:cwss}
 \|{{Z}}_k \| \leq    Q  \mu_E \frac{s}{P_x}  \max \{ r_E-1, 1\}     \triangleq \bar{\mu}_I^u, \quad \forall k.
\end{align}
For $\bar{\varrho}_I^u$, we note that $\| \sum_{k=1}^{\nts} W_{k}^2  \|  \leq (s/ N)^2$ and $\expectation[ \bar{p}_k^2 ] \leq  Q {\varrho_E}(\frac{N}{P_x})^2 $. Hence using \eqref{eqn:varbound1}, we have 
\begin{align} \label{eqn:varbound:cwss}
 \| \sum_{k=1}^{\nts} \expectation[ Z_k^2]  \| \leq   \varrho_E Q  (\frac{s}{P_x})^2    \triangleq  \bar{\varrho}_I^u.
\end{align}
Using \eqref{eqn:normbound:cwss}, \eqref{eqn:varbound:cwss} and the Matrix Bernstein Inequality shows that 
$
\| \sum_{k=1}^{\nts} Z_k \|   < \bar{r}^u
$
holds with probability greater than $p_{bt} = 1-f_{bt}(\bar{\mu}_I^u, \bar{\varrho}_I^u,\bar{r}^u)$.  
Therefore, we have the following
\begin{align}
\lambda_{min}({U_s}^\dagger G U_s)
&> \lambda_{min}(U_s^\dagger \bar{G} U_s )-\bar{r}^u\\
\label{eq:eigdiff1:cwss}
&=  \frac{s}{N}\lambda_{min}({F^s}^\dagger  \diag( \expectation[p_{k}] ) F^s)-\bar{r}^u\\
\label{eq:eigdiff2:cwss}
&= \mu_E Q \frac{s}{P_x} -\bar{r}^u
\end{align}
where \eqref{eq:eigdiff1:cwss} follows from \eqref{eqn:dftsampled} and \eqref{eq:eigdiff2:cwss}
 follows from the fact that  $\expectation[p_{k}] =\mu_E Q N/P_x$ and  ${F^s}^\dagger  F^s = I_s$. Now rescaling $\bar{\mu}_I^u$, $\bar{\varrho}_I^u$, $\bar{r}^u$ with $c$, $c^2$ and $c$, respectively where $c=Q \frac{s}{P_x}$ concludes the proof.

\section{Proof of  Lemma~\protect{\ref{offline:equidistantcwss}}}\label{sec:pf:offline:equidistantcwss}
The result is a consequence of \cite[Lemma 3.6]{ayca_eusipco2016} which considers the setting in 
Lemma~\protect{\ref{offline:equidistantcwss}} under off-line energy harvesting constraints. In particular, 
 \cite[Lemma 3.6]{ayca_eusipco2016} states that the most majorized solution under given energy constraints is optimal. In the setting of Lemma~\protect{\ref{offline:equidistantcwss}}, there is only one energy constraint which is a constraint on the total energy, i.e. sum of $p_k$.  Under  a total sum constraint, the most majorized allocation is  uniform allocation \cite[Ch.3]{b_marshallOlkin}, i.e. $p_i =\frac{E_{tot}}{s}\frac{N}{P_x}$. This concludes the proof.
 
Due to the slightly different setting of \cite[Lemma 3.6]{ayca_eusipco2016},  we also provide a step-by-step proof of  Lemma~\protect{\ref{offline:equidistantcwss}} below for the sake of clarity: 
Due to \eqref{eqn:dftsampled}, we have  $F_\Omega^N B_u F_\Omega^N= \frac{s}{N} F^s \diag(p_k) F^s$. Hence the error is given by 
\begin{align}
 \err &=  \sum_{i=1}^{s}  \frac{1}{1 + \frac{1}{\sigma_w^2}\frac{P_x}{s} \lambda_{i}(U_s^\herm G  U_s)} P_x \\
 &= \sum_{i=1}^{s}  \frac{1}{1 + \frac{1}{\sigma_w^2}\frac{P_x}{N} \lambda_{i}(F_s^\herm \diag(p_i)   F_s)} P_x \\
    &= \sum_{i=1}^{s}  \frac{1}{1 + \frac{1}{\sigma_w^2}\frac{P_x}{N} p_i} P_x
\end{align}
where we have used the fact that $F_s$ is unitary and  $\lambda_{i}(F_s^\herm \diag(p_k)   F_s) = p_i $. 
 Now $\err$ is a symmetric and Schur-convex function of $p_i$, $i=1,\ldots, N$ \cite[Ch.3]{b_marshallOlkin}. 
Hence an optimal solution is given by uniform $p_i$ \cite[Ch.3]{b_marshallOlkin}. 

%% file: a_EH_intermittent_online_JRNL.bbl
\begin{thebibliography}{10}

\bibitem{gunduzStamatiouMichelusiZorzi_2014}
D.~G\"{u}nd\"{u}z, K.~Stamatiou, N.~Michelusi, and M.~Zorzi, ``Designing
  intelligent energy harvesting communication systems,'' {\em IEEE
  Communications Magazine}, vol.~52, no.~1, pp.~210--216, 2014.

\bibitem{gilbert_comparison_2008}
J.~M. Gilbert and F.~Balouchi, ``Comparison of energy harvesting systems for
  wireless sensor networks,'' {\em International Journal of Automation and
  Computing}, vol.~5, pp.~334--347, Oct. 2008.

\bibitem{vullers_2010}
R.~Vullers, R.~Schaijk, H.~Visser, J.~Penders, and C.~Hoof, ``Energy
  {Harvesting} for {Autonomous} {Wireless} {Sensor} {Networks},'' {\em IEEE
  Solid-State Circuits Magazine}, vol.~2, no.~2, pp.~29--38, 2010.

\bibitem{visserVullers_2013}
H.~J. Visser and R.~J.~M. Vullers, ``{RF} {Energy} {Harvesting} and {Transport}
  for {Wireless} {Sensor} {Network} {Applications}: {Principles} and
  {Requirements},'' {\em Proceedings of the IEEE}, vol.~101, pp.~1410--1423,
  June 2013.

\bibitem{gorlatovaWallwaterZussman_2013}
M.~Gorlatova, A.~Wallwater, and G.~Zussman, ``Networking low-power energy
  harvesting devices: {Measurements} and algorithms,'' {\em IEEE Transactions
  on Mobile Computing}, vol.~12, no.~9, pp.~1853--1865, 2013.

\bibitem{ZhangHo_2013}
R.~Zhang and C.~K. Ho, ``{MIMO} broadcasting for simultaneous wireless
  information and power transfer,'' {\em {IEEE} Trans. Wireless Commun.},
  vol.~12, pp.~1989--2001, May 2013.

\bibitem{huangLarsson_2013}
K.~Huang and E.~Larsson, ``Simultaneous information and power transfer for
  broadband wireless systems,'' {\em {IEEE} Trans. Signal Process.},
  pp.~5972--5986, Dec. 2013.

\bibitem{ParkClerckx_2014}
J.~Park and B.~Clerckx, ``Joint wireless information and energy transfer in a
  ${K}$-user {MIMO} interference channel,'' {\em {IEEE} Trans. Wireless
  Commun.}, vol.~13, pp.~5781--5796, Oct. 2014.

\bibitem{ozcelikkaleDuman_2015_TWC}
A.~Ozcelikkale and T.~M. Duman, ``Linear precoder design for simultaneous
  information and energy transfer over two-user {MIMO} interference channels,''
  {\em {IEEE} Trans. Wireless Commun.}, vol.~14, pp.~5836--5847, Oct 2015.

\bibitem{OzelTutuncuogluYangUlukusYener_2011}
O.~Ozel, K.~Tutuncuoglu, J.~Yang, S.~Ulukus, and A.~Yener, ``Transmission with
  {Energy} {Harvesting} {Nodes} in {Fading} {Wireless} {Channels}: {Optimal}
  {Policies},'' {\em {IEEE} J. Sel. Areas Commun.}, vol.~29, pp.~1732--1743,
  Sept. 2011.

\bibitem{TutuncuogluYener_2012}
K.~Tutuncuoglu and A.~Yener, ``Optimum transmission policies for battery
  limited energy harvesting nodes,'' {\em {IEEE} Trans. Wireless Commun.},
  vol.~11, pp.~1180--1189, March 2012.

\bibitem{antepliUysalErkal_2011}
M.~A. Antepli, E.~Uysal-Biyikoglu, and H.~Erkal, ``Optimal {Packet}
  {Scheduling} on an {Energy} {Harvesting} {Broadcast} {Link},'' {\em {IEEE} J.
  Sel. Areas Commun.}, vol.~29, pp.~1721--1731, Sept. 2011.

\bibitem{yangUlukus_2012mac}
J.~Yang and S.~Ulukus, ``Optimal packet scheduling in a multiple access channel
  with energy harvesting transmitters,'' {\em Journal of Communications and
  Networks,}, vol.~14, no.~2, pp.~140--150, 2012.

\bibitem{ozelUlukus_2012}
O.~Ozel and S.~Ulukus, ``Achieving {AWGN} capacity under stochastic energy
  harvesting,'' {\em {IEEE} Trans. Inf. Theory}, vol.~58, pp.~6471--6483, Oct
  2012.

\bibitem{DongFarniaOzgur_2015}
Y.~Dong, F.~Farnia, and A.~\"{O}zg\"{u}r, ``Near optimal energy control and
  approximate capacity of energy harvesting communication,'' {\em {IEEE} J.
  Sel. Areas Commun.}, vol.~33, pp.~540--557, March 2015.

\bibitem{SrivastavaKoksal_2013}
R.~Srivastava and C.~E. Koksal, ``Basic {Performance} {Limits} and {Tradeoffs}
  in {Energy}-{Harvesting} {Sensor} {Nodes} {With} {Finite} {Data} and {Energy}
  {Storage},'' {\em IEEE/ACM Trans. on Networking}, vol.~21, pp.~1049--1062,
  Aug. 2013.

\bibitem{kazerouniOzgur_2015}
A.~Kazerouni and A.~Ozgur, ``Optimal online strategies for an energy harvesting
  system with {Bernoulli} energy recharges,'' in {\em 2015 {Inter.} {Symp.} on
  Modeling and {Opt.} in {Mobile}, {Ad} {Hoc}, and {Wireless} {Networks}
  ({WiOpt}),}, pp.~235--242, 2015.

\bibitem{NayyarBasarTeneketzisVeeravalli_2013}
A.~Nayyar, T.~Ba\c{s}ar, D.~Teneketzis, and V.~Veeravalli, ``Optimal strategies
  for communication and remote estimation with an energy harvesting sensor,''
  {\em {IEEE} Trans. Autom. Control}, vol.~58, pp.~2246--2260, Sept 2013.

\bibitem{blascoGunduzDohler_2013}
P.~Blasco, D.~Gunduz, and M.~Dohler, ``A {Learning} {Theoretic} {Approach} to
  {Energy} {Harvesting} {Communication} {System} {Optimization},'' {\em {IEEE}
  Trans. Commun.}, vol.~12, pp.~1872--1882, Apr. 2013.

\bibitem{ayca_unitaryIT2014}
A.~\"{O}z\c{c}elikkale, S.~Y\"{u}ksel, and H.~Ozaktas, ``Unitary precoding and
  basis dependency of {MMSE} performance for {G}aussian erasure channels,''
  {\em {IEEE} Trans. Inf. Theory}, vol.~60, pp.~7186--7203, Nov 2014.

\bibitem{nourianDeyAhlen_2015}
M.~Nourian, S.~Dey, and A.~Ahlen, ``Distortion {Minimization} in
  {Multi}-{Sensor} {Estimation} {With} {Energy} {Harvesting},'' {\em {IEEE} J.
  Sel. Areas Commun.}, vol.~33, pp.~524--539, Mar. 2015.

\bibitem{knornDeyAhlenQuevedo_2015}
S.~Knorn, S.~Dey, A.~Ahlen, and D.~E. Quevedo, ``Distortion {Minimization} in
  {Multi}-{Sensor} {Estimation} {Using} {Energy} {Harvesting} and {Energy}
  {Sharing},'' {\em {IEEE} Trans. Signal Process.}, vol.~63, pp.~2848--2863,
  June 2015.

\bibitem{NourianLeongDey_2014}
M.~Nourian, A.~Leong, and S.~Dey, ``Optimal energy allocation for {Kalman}
  filtering over packet dropping links with imperfect acknowledgments and
  energy harvesting constraints,'' {\em {IEEE} Trans. Autom. Control}, vol.~59,
  pp.~2128--2143, Aug 2014.

\bibitem{calvofullanaMatamorosAntonHaro_2015}
M.~Calvo-Fullana, J.~Matamoros, and C.~Anton-Haro, ``Reconstruction of
  {Correlated} {Sources} with {Energy} {Harvesting} {Constraints},'' in {\em
  European {Wireless} Conf. 2015}, pp.~1--6, 2015.

\bibitem{ayca_eusipco2016}
A.~\"{O}z\c{c}elikkale, T.~McKelvey, and M.~Viberg, ``Transmission strategies
  for remote estimation under energy harvesting constraints,'' in {\em European
  Signal Process. Conf. ({EUSIPCO})}, pp.~572 -- 576, 2016.

\bibitem{GangulaGunduzGesbert_2015}
R.~Gangula, D.~G\"und\"uz, and D.~Gesbert, ``Distributed compression and
  transmission with energy harvesting sensors,'' in {\em 2015 IEEE
  International Symposium on Information Theory (ISIT)}, pp.~1139--1143, 2015.

\bibitem{zhaoChenZhang_2015}
Y.~Zhao, B.~Chen, and R.~Zhang, ``Optimal power management for remote
  estimation with an energy harvesting sensor,'' {\em {IEEE} Trans. Wireless
  Commun.}, vol.~14, pp.~6471--6480, Nov. 2015.

\bibitem{OrhanGunduzErkip_2015}
O.~Orhan, D.~G\"und\"uz, and E.~Erkip, ``Source-channel coding under energy,
  delay, and buffer constraints,'' {\em {IEEE} Trans. Wireless Commun.},
  vol.~14, pp.~3836--3849, July 2015.

\bibitem{GastparRimoldiVetterli_2003}
M.~Gastpar, B.~Rimoldi, and M.~Vetterli, ``To code, or not to code: lossy
  source-channel communication revisited,'' {\em IEEE Transactions on
  Information Theory}, vol.~49, pp.~1147--1158, May 2003.

\bibitem{gastpar_2008}
M.~Gastpar, ``Uncoded transmission is exactly optimal for a simple {Gaussian}
  sensor network,'' {\em {IEEE} Trans. Inf. Theory}, vol.~54, no.~11,
  pp.~5247--5251, 2008.

\bibitem{ayca_isit2016}
A.~\"{O}z\c{c}elikkale, T.~McKelvey, and M.~Viberg, ``Performance bounds for
  remote estimation with an energy harvesting sensor,'' in {\em {IEEE} Int.
  Symp. Information Theory (ISIT)}, pp.~460--464, 2016.

\bibitem{neeser_proper_1993}
F.~D. Neeser and J.~L. Massey, ``Proper complex random processes with
  applications to information theory,'' {\em {IEEE} Trans. Inf. Theory},
  vol.~39, no.~4, pp.~1293--1302, 1993.

\bibitem{GrayToeplitzReview}
R.~M. Gray, {\em Toeplitz and Circulant Matrices: a Review}.
\newblock Now Publishers Inc., 2006.

\bibitem{TulinoCaireVerduShamai_2013}
A.~Tulino, G.~Caire, S.~Verdu, and S.~Shamai, ``Support recovery with sparsely
  sampled free random matrices,'' {\em {IEEE} Trans. Inf. Theory}, vol.~59,
  no.~7, pp.~4243--4271, 2013.

\bibitem{tirronen_2013}
T.~Tirronen, A.~Larmo, J.~Sachs, B.~Lindoff, and N.~Wiberg,
  ``Machine-to-machine communication with long-term evolution with reduced
  device energy consumption,'' {\em Trans. on Emerging Telecomm. Tech.},
  vol.~24, pp.~413--426, June 2013.

\bibitem{BahceciKhandani_2008}
I.~Bahceci and A.~Khandani, ``Linear estimation of correlated data in wireless
  sensor networks with optimum power allocation and analog modulation,'' {\em
  {IEEE} Trans. Commun.}, vol.~56, pp.~1146--1156, July 2008.

\bibitem{LeePetersen_1976}
K.-H. Lee and D.~Petersen, ``Optimal linear coding for vector channels,'' {\em
  {IEEE} Trans. Commun.}, vol.~24, pp.~1283--1290, Dec 1976.

\bibitem{TanGunduzVillardebo_2016}
O.~Tan, D.~G\"und\"uz, and J.~G\'omez-Vilardeb\'o, ``Linear transmission of
  composite {G}aussian measurements over a fading channel under delay
  constraints,'' {\em {IEEE} Trans. Wireless Commun.}, vol.~15, pp.~4335--4347,
  June 2016.

\bibitem{b_andersonMoore_optFiltering}
B.~D.~O. Anderson and J.~B. Moore, {\em Optimal filtering}.
\newblock Prentice-Hall, 1979.

\bibitem{foucartRauhut_2013}
S.~Foucart and H.~Rauhut, {\em A {Mathematical} {Introduction} to {Compressive}
  {Sensing}}.
\newblock Springer, 2013.

\bibitem{Papoulis_1991book}
A.~Papoulis, {\em Probability, Random Variables, and Stochastic Processes}.
\newblock Mcgraw-Hill, 3rd~ed., 1991.

\bibitem{b_HornJohnson_TMA}
R.~A. Horn and C.~R. Johnson, {\em Topics in Matrix Analysis}.
\newblock Cambridge University Press, 1991.

\bibitem{tropp_expected_2015}
J.~A. Tropp, ``The expected norm of a sum of independent random matrices: {An}
  elementary approach,'' {\em arXiv preprint}, 2015.

\bibitem{b_marshallOlkin}
A.~W. Marshall and I.~Olkin, {\em Inequalities: Theory of Majorization and its
  Applications}.
\newblock Academic Press, 1979.

\end{thebibliography}
